\documentclass[prd,aps,10pt,notitlepage,showpacs,nofootinbib,tightenlines]{revtex4}
\usepackage{mathrsfs}
\usepackage{amsmath}
\usepackage{amssymb}
\usepackage{epsfig}
\usepackage{graphicx}
\usepackage{booktabs}
\usepackage{multirow}
\usepackage{subfigure}
\usepackage{bm}
\usepackage{times}
\usepackage{braket}
\usepackage{color}
\usepackage{slashed}
\usepackage{hyperref}
\usepackage{threeparttable}
\newcommand{\beq}{\begin{eqnarray}}
\newcommand{\eeq}{\end{eqnarray}}
\newcommand{\non}{\nonumber\\ }

\definecolor{Red}{rgb}{1.,0.,0.}

\definecolor{Blue}{rgb}{0.,0.,1.}

\definecolor{nicered}{rgb}{0.7,0.1,0.1}
\definecolor{nicegreen}{rgb}{0.1,0.5,0.1}
\hypersetup{colorlinks,citecolor=nicegreen,linkcolor=nicered}

\def \epjc{ Eur. Phys. J. C }

\def \jpg{  J. Phys. G }
\def \npb{  Nucl. Phys. B }
\def \plb{  Phys. Lett. B }

\def \prd{  Phys. Rev. D }
\def \prl{  Phys. Rev. Lett.  }

\def \jhep{ J. High Energy Phys. }

\begin{document}

\title{$CP$-violating observables of four-body $B_{(s)} \to (\pi\pi)(K\bar{K})$ decays in perturbative QCD}

\author{Da-Cheng Yan$^1$}         
\author{Yan Yan$^{1}$}    
\author{Zhou Rui$^2$}              \email[Corresponding author:]{jindui1127@126.com}
\affiliation{$^1$ School of Mathematics and Physics, Changzhou University, Changzhou, Jiangsu 213164, China}
\affiliation{$^2$ College of Sciences, North China University of Science and Technology, Tangshan, Hebei 063210, China}
\date{\today}

\begin{abstract}
In this work, we investigate six helicity amplitudes of the four-body $B_{(s)} \to (\pi\pi)(K\bar{K})$ decays via an angular analysis in the perturbative QCD (PQCD) approach.
The $\pi\pi$ invariant mass spectrum is dominated by the vector resonance $\rho(770)$ together with scalar resonance $f_0(980)$, while the vector resonance $\phi(1020)$
and scalar resonance $f_0(980)$ are expected to contribute in the $K\bar{K}$ invariant mass range.
We extract the two-body branching ratios ${\cal B}(B_{(s)}\to \rho\phi)$ from the corresponding four-body decays
$B_{(s)}\to \rho\phi\to (\pi\pi)(K \bar K)$ based on the narrow width approximation.
The predicted ${\cal B}(B^0_{s}\to \rho\phi)$ agrees well with the current experimental data within errors.
The longitudinal polarization fractions of the $B_{(s)}\to \rho\phi$ decays are found to be as large as $90\%$,
basically consistent with the previous two-body predictions within uncertainties.
In addition to the direct $CP$ asymmetries, the triple-product asymmetries (TPAs) originating from the interference among various helicity amplitudes are also presented for the first time.
Since the $B_s^0\to \rho^0\phi\to(\pi^+\pi^-)(K^+K^-)$ decay is induced by both tree and penguin operators,
the values of the ${\cal A}^{\rm CP}_{\rm dir}$ and  ${\cal A}^{1}_{\text{T-true}}$
are calculated to be $(21.8^{+2.7}_{-3.3})\%$ and $(-10.23^{+1.73}_{-1.56})\%$  respectively.
While for pure penguin decays $B^0\to \rho^0\phi\to(\pi^+\pi^-)(K^+K^-)$ and $B^+\to \rho^+\phi\to(\pi^+\pi^0)(K^+K^-)$,
both the direct $CP$ asymmetries and ``true" TPAs are naturally expected to be zero in the standard model (SM) due to the absence of the weak phase difference.
The ``fake" TPAs requiring no weak phase difference are usually none zero for all considered decay channels.
The sizable ``fake" ${\cal A}^{1}_{\text{T-fake}}=(-20.92^{+6.26}_{-2.80})\%$ of the
$B^0\to \rho^0\phi\to(\pi^+\pi^-)(K^+K^-)$ decay
is predicted in the PQCD approach,
which provides valuable information on the final-state interactions.
The above predictions can be tested by the future LHCb and Belle-II experiments.

\end{abstract}

\maketitle

\section{Introduction}
In the past several years, the study of charmless nonleptonic decays of $B$ meson has evoked considerable experimental and theoretical interest, primarily because of the importance of these processes in understanding the phenomenon of $CP$ violation.
The decay amplitude for ``tree-level" $b\to u$ transition is much smaller than the one for dominant $b\to c$ transition due to the ratio of Cabibbo-Kobayashi-Maskawa (CKM) matrix elements $|V_{ub}|^2/|V_{cb}|^2\approx 10^{-2}$.
Transitions to $s$ and $d$ quarks are effective flavor-changing neutral currents proceeding mainly by one-loop ``penguin" amplitudes, and are also suppressed.
The flavor-changing neutral current decay modes provide a sensitive probe for the effect of physics beyond the SM, since their amplitudes are dominant by the penguin diagrams.
The understanding of the relative importance of tree and penguin amplitudes will be crucial in studies of $CP$ asymmetries in $B$ meson decays.
A non-vanishing direct $CP$ violation needs the interference of at least two amplitudes with a weak phase difference $\Delta \phi$ and a strong phase difference $\Delta \delta$.
The direct $CP$ violation is proportional to $\sin\Delta \phi \sin \Delta \delta$.
The key point is that the direct $CP$ violation can only be produced when there is a nonzero strong phase difference.
Hence, if the strong phases are quite small, the magnitude of the direct $CP$ violation is close to zero.
In this case, there is another class of $CP$-violating effects which has triggered less attention so far and can reveal the presence of new physics: triple product asymmetries (TPAs).
A scalar triple product takes a generic form
$\vec{v}_1\cdot (\vec{v}_2 \times \vec{v}_3)$,
where each $\vec{v}_i$ is a spin or momentum of the final-state particle.
The TPAs are odd under time reversal ($T$) and  also contribute potential signals of $CP$ violation by the $CPT$ theorem.
These TPAs go as $\sin\Delta \phi \cos \Delta \delta$, which provide useful complementary information on direct $CP$ violation.
Even in the absence of $CP$ violation effects, $T$-odd triple products (also called ``fake" TPAs), which are proportional to $\cos\Delta \phi \sin \Delta \delta$, can provide further insight on new physics since most TPAs are expected to be tiny within the SM~\cite{plb701-357}.

\begin{figure}[tbp]
	\centering
	\includegraphics[scale=0.7]{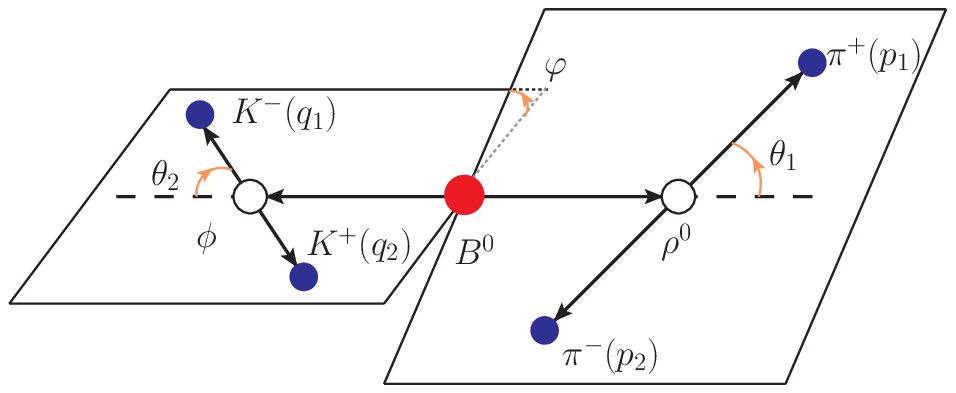}
\vspace{-14cm}
	\caption{Graphical definitions of the helicity angles $\theta_1$, $\theta_2$ and $\varphi$ for the $B^0 \to \phi\rho^0$ decay, with each quasi-two-body intermediate resonance decaying to two pseudoscalars ($\rho^0 \to \pi^+\pi^-$ and $\phi \to K^+K^-$).
	$\theta_{1,2}$ is denoted as the angle between the direction of motion of $K^-$ or $\pi^+$ in the $\phi$ or $\rho^0$ rest frame and $\phi$ or $\rho^0$ in the $B^0$ rest frame, and $\varphi$ is the angle between the plane defined by $K^+K^-$ and the plane defined by $\pi^+\pi^-$ in the $B^0$ rest frame.}
\label{fig1}
\end{figure}

A nontrivial triple product requires at least four particles in the final state.
$B_{(s)} \to VV,VS,SV,SS$ decays are usually treated as two-body final states on the theoretical side and have been studied in the two-body framework using various theoretical approaches such as QCD factorization (QCDF)~\cite{npb774-64,prd77-014034,prd80-114008,prd80-114026,1206-4106,prd87-114001,2202.08073}, PQCD approaches
~\cite{plb622-63,prd72-054015,jpg32-101,prd73-014011,prd74-114010,prd73-104024,prd76-074018,prd81-074014,epjc67-163,prd82-034036,prd88-094003,prd91-054033,npb935-17,epjc82-59,prd102-116007,epjc82-177}, the soft-collinear-effective theory (SCET)~\cite{prd70-054015,prd74-034010,npb692-232,prd72-098501,prd72-098502,prd96-073004} and the factorization-assisted topological amplitude approach (FAT)~\cite{epjc77-333}.
While they are at least four-body decays on the experimental side shown in Fig.~\ref{fig1}, since the meson $V=\rho,\phi$ is a vector resonance and $S=f_0(980)$ is a scalar resonance with a sizable branching fraction into two pseudoscalar mesons, respectively
\footnote{ For the sake of simplicity, we
generally use the abbreviation $f_0=f_0(980)$, $\rho^0=\rho(770)^0$, $\phi=\phi(1020)$ in the following sections. }.
The $B$ decays to $VV$ are complicated by the presence of one amplitude with longitudinal polarization $A_0$ and two amplitudes with transverse polarization $A_{\parallel}$ and $A_{\perp}$, which is parallel
or perpendicular to each other, respectively.
The first two states $A_0$ and $A_{\parallel}$ are $CP$ even, while the last one $A_{\perp}$ is $CP$ odd.
Interference between the $CP$-even ($A_0$, $A_{\parallel}$) and $CP$-odd ($A_{\perp}$) amplitudes can generate TPAs in angular distributions, which may signal unexpected $CP$ violation due to physics beyond the SM.
Recently, TPAs have already been measured by Belle, {\it BABAR}, CDF and LHCb Collaborations~\cite{prl95-091601,prd76-031102,prl107-261802,plb713-369,LHCb:2013xyz,LHCb:2014xzf,prd90-052011,jhep07-166,jhep05-026,LHCb:2019jgw}.
Phenomenological investigations on TPAs have been conducted intensively in the
literature~\cite{prd39-3339,npbps13-487,ijmpa19-2505,prd84-096013,plb701-357,prd86-076011,prd88-016007,prd92-076013,prd87-116005,2211.07332}.

\begin{figure}[tbp]
	\centering
	\includegraphics[scale=0.8]{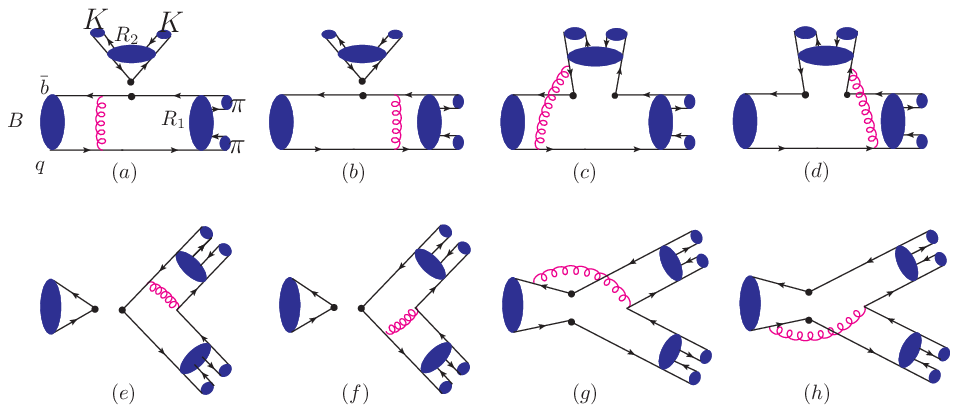}
\vspace{-16cm}
	\caption{Typical leading-order Feynman diagrams for the four-body decays $B \to R_1R_2\to (\pi\pi)(K \bar K)$ with $q=(d,s)$, where the symbol $\bullet$ denotes a weak interaction vertex. The diagrams ($a$)-($d$) represent the $B \to (R_1\to) \pi\pi$ transition, as well as the diagrams ($e$)-($h$) for annihilation contributions.
If we exchange the position of $R_1(\to \pi\pi)$ and $R_2 (\to K\bar K)$, we will find the diagrams ($a$)-($d$) for the $B \to (R_2\to) K\bar K$ transition.}
\label{fig2}
\end{figure}

In this work, we study the four-body decays $B_{(s)} \to (\pi\pi)(K \bar K)$ in the PQCD approach based on $k_T$ factorization with the relevant Feynman diagrams illustrated
in Fig.~\ref{fig2}.
For a comparison with the LHCb experiment~\cite{prd95-012006}, the invariant mass of the $\pi\pi$ pair ranges from 400 MeV to 1600 MeV and the invariant mass for $K \bar K$ pair is restricted to be within $\pm30 {\rm MeV}$ of the known mass of the $\phi$ meson.
The $\pi\pi$ spectrum is dominated by the vector $
\rho$ resonance and the scalar resonance $f_0$.
In the considered $K \bar K$ invariant-mass range, the vector resonance $\phi$ is expected to contribute, together with the scalar resonance $f_0$.
In addition to the branching fractions, the fraction of a given polarization state is an interesting observable and investigated in this work, as well as other observables constructed from the helicity amplitudes like TPAs.
As is known, the longitudinal polarization should dominate based on the quark helicity analysis in the factorization assumption ~\cite{zpc1-269,prd64-117503}.
In sharp contrast to these expectations, large transverse polarization (around 50$\%$) is observed in $B\to K^*\phi$, $B\to K^*\rho$ and $B_s\to \phi\phi$ decays~\cite{prl91-201801,prd78-092008,prd85-072005,LHCb:2014xzf,LHCb:2019jgw}, which poses an interesting challenge for the theory.

It should be stressed that four-body decay is still at its early stage from the theoretical point of view since the factorization formalism that describes a multi-body decay in full phase space is not yet available at present.
As a first step, we can only restrict ourselves to the specific kinematical configurations in which each two particles fly collinearly and two pairs of final state particles recoil back in the rest frame of the $B$ meson, see Fig.~\ref{fig1}.
Then the dynamics associated with the pair of final state mesons can be factorized into a two-meson distribution amplitude (DA)
$\Phi_{h_1h_2}$~\cite{MP,MT01,MT02,MT03,NPB555-231,Grozin01,Grozin02}.
Thereby, the typical PQCD factorization formula for the considered four-body decay amplitude can be described,
\begin{eqnarray}\label{amplitude}
\mathcal{A}=\Phi_B\otimes H\otimes \Phi_{KK}\otimes\Phi_{\pi\pi},
\end{eqnarray}
where $\Phi_B$ is the universal wave function of the $B$ meson and absorbs the non-perturbative dynamics in the process.
The $\Phi_{KK}$ ($\Phi_{\pi\pi}$) is the two-hadron DA, which involves the resonant and nonresonant interactions between the two moving collinearly mesons.
The hard kernel $H$ describes the dynamics of the strong and electroweak interactions in four-body hadronic decays in a similar way as the one for the corresponding two-body decays.
The $S$ and $P$-wave contributions are parametrized into the corresponding timelike form factors involved in the two-meson DAs, whose
normalization form factors are assumed to take the Flatt\'e model~\cite{plb63-228} for $f_0$, and relativistic Breit-Wigner (BW) function for $\phi$~\cite{BW-model}
and the Gounaris-Sakurai (GS)
model~\cite{prl21-244} for $\rho$.
An important breakthrough in the theory of four-body $B$ meson decays has been achieved based on the quasi-two-body-decay mechanism.
Recently, the localized $CP$ violation and branching fraction of the four-body decay
$\bar{B}^0\to K^-\pi^+\pi^+\pi^-$ have been calculated by employing a quasi-two-body QCDF
approach in Refs.~\cite{1912-11874,2008-08458}.
In our previous works~\cite{zjhep,Li:2021qiw,prd105-053002,prd105-093001,epjc83-974}, the PQCD factorization formalism
based on the quasi-two-body-decay mechanism for four-body $B$ meson decays has been well established.
Within the framework of PQCD approach, the branching ratios and direct $CP$ asymmetries of four-body decays $B_s^0 \to\pi\pi\pi\pi$ have also been studied~\cite{Liang:2022mrz}.

The layout of the present paper is organized as follows.
In Sec.~\ref{sec:2}, we give a brief introduction for the triple product asymmetries analyzed in our work.
The kinematics and the formalism of PQCD on four body decays are presented in Sec.~\ref{sec:3}.
The numerical values and some discussions will be given in Sec.~\ref{sec:4}.
Section~\ref{sec:5} contains our conclusions.
The APPENDIX~\ref{samp} and \ref{TMDAs} collect the $S$-wave decay amplitudes and the two-meson DAs adopted in our calculations respectively.

\section{$CP$ violating observables}\label{sec:2}
\subsection{Angular distribution and Helicity amplitudes}
Taking the four-body decay $B^0\to \rho^0\phi\to (\pi^+\pi^-)(K^+K^-)$ depicted in Fig.~\ref{fig1} as an example,
the study of the angular distribution usually employs three helicity angles: $\theta_{1}$, $\theta_{2}$ and $\varphi$.
We denote $\theta_1(\theta_2)$ as the polar angle between the $\pi^+$($K^-$) direction in the $\pi^+\pi^-$($K^+K^-$)
rest frame and the $\pi^+\pi^-$($K^+K^-$) direction in the $B^0$ rest frame.
The angle between the planes defined by $\pi^+\pi^-$ and $K^+K^-$ pairs in the $B^0$ rest frame will be denoted by $\varphi$.

In the four-body decays $B_{(s)}\rightarrow (\pi\pi)(K{\bar K})$,
the final state meson pairs $\pi\pi$ and $K{\bar K}$
can be produced in both $S$ and $P$-wave configurations in the selected invariant mass regions.
The total decay amplitudes shall involve six helicity components $A_h$ with $h=VV(3), VS, SV$,  and $SS$,
where $V$ and $S$ denote the  vector and scalar resonances respectively.
The first three, commonly referred to as the $P$-wave amplitudes, are associated with
the final states where both $\pi\pi$ and $K{\bar K}$ pairs come from intermediate vector mesons.
Following the definitions given in Refs.~\cite{ zjhep,Li:2021qiw},
the $P$-wave decay amplitudes can be decomposed into three components in the transversity basis,
including longitudinal $A_0$, parallel $A_{\parallel}$, and perpendicular $A_\perp$.
As the $S$-wave $\pi\pi$ and $K {\bar K}$ pair can arise from the intermediate resonances $R_1$ and $R_2$ labelled in Fig.~\ref{fig2}(a),
the related two single $S$-wave decay amplitudes are described as $A_{SV}$ and $A_{VS}$ respectively,
which are physical different.
The double $S$-wave amplitude $A_{SS}$ is associated with the final states,
where both $\pi\pi$ and $K{\bar K}$ meson pairs are generated in the $S$ wave.
All of the above mentioned helicity amplitudes of the four-body decays $B_{(s)}\rightarrow (\pi\pi)(K{\bar K})$ are summarized as follows:
\begin{eqnarray}
A_{VV}&:& B_{(s)} \rightarrow \rho(\to\pi\pi) \phi (\to K{\bar K}), \nonumber\\
A_{VS}&:& B_{(s)} \rightarrow \rho(\to\pi\pi) f_0 (\to K{\bar K}),\nonumber\\
A_{SV}&:& B_{(s)} \rightarrow f_0(\to\pi\pi) \phi (\to K{\bar K}) ,\nonumber\\
A_{SS}&:& B_{(s)} \rightarrow f_0(\to\pi\pi) f_0 (\to K{\bar K}) .
\end{eqnarray}

\subsection{Triple-product asymmetries in four-body $B_{(s)} \to (\pi\pi)(K {\bar K})$ decays}
As stressed in the Introduction, TPAs and direct $CP$ violation can complement each other.
TPA is another class of $CP$-violating effect, which has received considerably less attention and can also reveal the presence of new physics.
In this section, we will briefly introduce the TPAs in the present work.

In the four-body decays $B_{(s)} \to (\pi\pi)(K {\bar K})$,
one can usually measure the four final state particles' momenta in the $B_{(s)}$ meson rest frame.
We define three unit vectors: $\hat{n}_{R_1}(\hat{n}_{R_2})$ perpendicular to the $R_1(R_2)$
decay plane and $\hat{z}$ in the direction of $R_1$ in the $B_{(s)}$ rest frame.
Thus we have
\begin{eqnarray}\label{TP1}
\hat{n}_{R_1}\cdot \hat{n}_{R_2}=\cos\varphi, ~\quad \hat{n}_{R_1}\times \hat{n}_{R_2}=\sin\varphi \hat{z},
\end{eqnarray}
implying the $T$-odd scalar triple products
\begin{eqnarray}\label{TP2}
(\hat{n}_{R_1}\times \hat{n}_{R_2})\cdot \hat{z}&=&\sin\varphi,\\
2(\hat{n}_{R_1}\cdot \hat{n}_{R_2})(\hat{n}_{R_1}\times \hat{n}_{R_2})\cdot\hat{z}&=&\sin2\varphi .
\end{eqnarray}
A $T$-odd asymmetry in the $B$ decay can usually be defined by an asymmetry between the number
of events  with positive and negative values of $\sin\varphi$ or $\sin2\varphi$,
\begin{eqnarray} \label{eq:ATs}
\mathcal{A}_{\text{T}}^1&=&\frac{\Gamma(\cos\theta_1\cos\theta_2\sin\varphi>0)-\Gamma(\cos\theta_1\cos\theta_2\sin\varphi<0)}
{\Gamma(\cos\theta_1\cos\theta_2\sin\varphi>0)+\Gamma(\cos\theta_1\cos\theta_2\sin\varphi<0)},\\
\mathcal{A}_{\text{T}}^2&=&\frac{\Gamma(\sin2\varphi>0)-\Gamma(\sin2\varphi<0)}
{\Gamma(\sin2\varphi>0)+\Gamma(\sin2\varphi<0)}.
\end{eqnarray}

In our calculations,
we will focus on the TPAs originating from the interference between the $CP$-odd amplitude $A_\perp$ and
the other two $CP$-even amplitudes $A_0$ and $A_{\parallel}$ in the $B_{(s)}\to \rho\phi\to (\pi\pi)(K {\bar K})$ decays,
which can be derived from the partially integrated differential decay rates as~\cite{prd84-096013,jhep07-166},
\begin{eqnarray} \label{eq:ATs2}
\mathcal{A}_{\text{T}}^1&=&\frac{\Gamma(\cos\theta_1\cos\theta_2\sin\varphi>0)-\Gamma(\cos\theta_1\cos\theta_2\sin\varphi<0)}
{\Gamma(\cos\theta_1\cos\theta_2\sin\varphi>0)+\Gamma(\cos\theta_1\cos\theta_2\sin\varphi<0)}\nonumber\\&=&
-\frac{2\sqrt{2}}{\pi \mathcal{D}}\int d\omega_1 d\omega_2k(\omega_1)k(\omega_2)k(\omega_1,\omega_2) \text{Im}[A_{\perp}A_0^*],\label{eq:AT1}\\
\mathcal{A}_{\text{T}}^2&=&\frac{\Gamma(\sin2\varphi>0)-\Gamma(\sin2\varphi<0)}
{\Gamma(\sin2\varphi>0)+\Gamma(\sin2\varphi<0)}\nonumber\\&=&
-\frac{4}{\pi \mathcal{D}}\int d\omega_1 d\omega_2k(\omega_1)k(\omega_2)k(\omega_1,\omega_2) \text{Im}[A_{\perp}A_{\parallel}^*],\label{eq:AT2}
\end{eqnarray}
with the denominator
\begin{equation}
\mathcal{D}=\int d\omega_1 d\omega_2k(\omega_1)k(\omega_2)k(\omega_1,\omega_2)(|A_0|^2+|A_{\parallel}|^2+|A_{\perp}|^2),
\end{equation}
 and the invariant mass of the final state meson pair $\omega_{1(2)}$.
The factor $k(\omega_1,\omega_2)$ represents  the magnitude of the three-momentum of the
meson pair in the $B_{(s)}$ meson rest frame,
\begin{eqnarray}
k(\omega_1,\omega_2)&=&\frac{\sqrt{[m_{B_{(s)}}^2-(\omega_1+\omega_2)^2][m_{B_{(s)}}^2-(\omega_1-\omega_2)^2]}}{2m_{B_{(s)}}},
\end{eqnarray}
where $m_{B_{(s)}}$ is the mass of the $B_{(s)}$ meson.
The kinematic variable $k(\omega)=\sqrt{\lambda(\omega^2,m_{h_1}^2,m_{h_2}^2)}/{(2\omega)}$
is defined in the $h_1h_2$ center-of-mass frame,
with the K$\ddot{a}$ll$\acute{e}$n function $\lambda (a,b,c)= a^2+b^2+c^2-2(ab+ac+bc)$ and $m_{h_1,h_2}$ being the final state mass.

It should be noted that
although  the two TPAs  given in Eqs.~(\ref{eq:AT1})
and~(\ref{eq:AT2}) in terms of transversity amplitudes are odd under time reversal,
they are not genuine $CP$ violation.
For example,
the integrands $\text{Im}(A_{\perp}A_{i}^*)$ $(i=0,\parallel)$ in the above TPAs can be expended in the form of
 $|A_{\perp}||A_{i}^*|\sin(\Delta\phi+\Delta \delta)$,
with $\Delta\phi$ and  $\Delta \delta$ representing the weak and strong phase differences between the two corresponding transversity amplitudes
$A_{\perp}$ and $A_{i}^*$.
The term $|A_{\perp}||A_{i}^*|\sin(\Delta\phi+\Delta \delta)$ can be nonzero even in the absence of any weak phases,
as long as the strong phase difference $\Delta \delta$ is nonzero.
Thus the two TPAs $\mathcal{A}_{\text{T}}^1$ and $\mathcal{A}_{\text{T}}^2$ can not reflect a true signal of $CP$ violation.
However, one can still obtain a true $CP$-violating asymmetry by comparing $\mathcal{A}_{\text{T}}$ with $ \mathcal{\bar A}_{\text{T}}$ ,
where $ \mathcal{\bar A}_{\text{T}}$ is the $T$-odd asymmetry measured in the $\bar B_{(s)}$ decay process.
We denote the helicity amplitudes  for the $CP$-conjugate decay process by ${\bar A}_{0}$, ${\bar A}_{\|}$ and ${\bar A}_{\bot}$,
which can be obtained by applying the following transformations:
\begin{eqnarray}
A_0\to \bar{A}_0, \quad A_{\parallel} \to \bar{A}_{\parallel}, \quad A_{\perp} \to -\bar{A}_{\perp}.
\end{eqnarray}
The TPAs  of the $\bar B_{(s)}$ decays can be defined similarly,
but with a multiplicative minus sign.
We then have the TPAs for the charge-averaged decay rates
\begin{eqnarray}
\mathcal{A}_\text{T-true}^{1,\text{ave}}&\equiv& \frac{[\Gamma(S_{1}>0)+\bar{\Gamma}(\bar{S}_{1}>0)]-[\Gamma(S_{1}<0)+\bar{\Gamma}(\bar{S}_{1}<0)]}
{[\Gamma(S_{1}>0)+\bar{\Gamma}(\bar{S}_{1}>0)]+[\Gamma(S_{1}<0)+\bar{\Gamma}(\bar{S}_{1}<0)]}\non
&=&-\frac{2\sqrt{2}}{\pi(\mathcal{D}+\bar{\mathcal{D}})}\int d\omega_1 d\omega_2k(\omega_1)k(\omega_2)k(\omega_1,\omega_2)
\text{Im}[A_{\perp}A_{0}^*-\bar{A}_{\perp}\bar{A}_{0}^*] , \label{tpatrueave1}\\
\mathcal{A}_\text{T-true}^{2,\text{ave}}&\equiv& \frac{[\Gamma(S_{2}>0)+\bar{\Gamma}(\bar{S}_{2}>0)]-[\Gamma(S_{2}<0)+\bar{\Gamma}(\bar{S}_{2}<0)]}
{[\Gamma(S_{2}>0)+\bar{\Gamma}(\bar{S}_{2}>0)]+[\Gamma(S_{2}<0)+\bar{\Gamma}(\bar{S}_{2}<0)]}\non
&=&-\frac{4}{\pi(\mathcal{D}+\bar{\mathcal{D}})}\int d\omega_1 d\omega_2k(\omega_1)k(\omega_2)k(\omega_1,\omega_2)
\text{Im}[A_{\perp}A_{\|}^*-\bar{A}_{\perp}\bar{A}_{\|}^*] , \label{tpatrueave2}\\
\mathcal{A}_\text{T-fake}^{1,\text{ave}}&\equiv& \frac{[\Gamma(S_{1}>0)-\bar{\Gamma}(\bar{S}_{1}>0)]-[\Gamma(S_{1}<0)-\bar{\Gamma}(\bar{S}_{1}<0)]}
{[\Gamma(S_{1}>0)+\bar{\Gamma}(\bar{S}_{1}>0)]+[\Gamma(S_{1}<0)+\bar{\Gamma}(\bar{S}_{1}<0)]}\non
&=&-\frac{2\sqrt{2}}{\pi(\mathcal{D}+\bar{\mathcal{D}})}\int d\omega_1 d\omega_2k(\omega_1)k(\omega_2)k(\omega_1,\omega_2)
\text{Im}[A_{\perp}A_{0}^*+\bar{A}_{\perp}\bar{A}_{0}^*] , \label{tpatrueave1}\\
\mathcal{A}_\text{T-fake}^{2,\text{ave}}&\equiv& \frac{[\Gamma(S_{2}>0)-\bar{\Gamma}(\bar{S}_{2}>0)]-[\Gamma(S_{2}<0)-\bar{\Gamma}(\bar{S}_{2}<0)]}
{[\Gamma(S_{2}>0)+\bar{\Gamma}(\bar{S}_{2}>0)]+[\Gamma(S_{2}<0)+\bar{\Gamma}(\bar{S}_{2}<0)]}\non
&=&-\frac{4}{\pi(\mathcal{D}+\bar{\mathcal{D}})}\int d\omega_1 d\omega_2k(\omega_1)k(\omega_2)k(\omega_1,\omega_2)
\text{Im}[A_{\perp}A_{\|}^*+\bar{A}_{\perp}\bar{A}_{\|}^*] , \label{tpatrueave2}
\end{eqnarray}
with $\bar\Gamma$ being the $CP$-conjugate decay rate,
the denominator
\begin{equation}
\mathcal{\bar D}=\int d\omega_1 d\omega_2k(\omega_1)k(\omega_2)k(\omega_1,\omega_2) (|\bar{A}_0|^2+|\bar{A}_{\parallel}|^2+|\bar{A}_{\perp}|^2),
\end{equation}
and the variables
\begin{eqnarray}
S_{1}&=&\cos\theta_1\cos\theta_2\sin\varphi,\quad S_2=\cos\theta_1\cos\theta_2\sin 2\varphi\non
{\bar S}_{1}&=&\cos{\bar\theta_1}\cos{\bar \theta_2}\sin{\bar \varphi},\quad
{\bar S}_{2}=\cos{\bar\theta_1}\cos{\bar \theta_2}\sin 2{\bar \varphi}.
\end{eqnarray}
It is shown that $\mathcal{A}_\text{T-true}^{1(2),\text{ave}}$ in terms of $\text{Im}[A_{\perp}A_{0(\|)}^*-\bar{A}_{\perp}\bar{A}_{0(\|)}^*]$
is proportional to $\sin\Delta\phi\cos\Delta\delta$.
They can be nonzero only in the presence of the weak phase difference $\Delta\phi$.
Therefore,
the ``true" averaged TPAs can provide extra measurements of $CP$ violation.
What's more,
compared with the direct $CP$ asymmetries ,
$\mathcal{A}_\text{T-true}^{1(2),\text{ave}}$
does not suffer the suppression from the strong phase difference,
which reaches the maximal value when the strong phase difference vanishes.
On the contrary, $\mathcal{A}_\text{T-fake}^{1(2),\text{ave}}\propto\cos\Delta\phi\sin\Delta\delta$ are not a  $CP$-violating signal as it is nonzero even in the absence of $CP$-violating phases.
Such a quantity will be referred as a ``fake" asymmetry ($CP$ conserving),
and simply reflects the effect of strong phases~\cite{prd84-096013,plb701-357}, instead of $CP$ violation.

In order to make a direct comparison with the future measurements,
we also calculate the so-called ``true"  and ``fake"  TPAs as follows,
\begin{eqnarray}
{\cal A}_{\text{T-true}}^{1}&=&\frac{1}{2}({\cal A}_{\text{T}}^{1}+\bar{{\cal A}}_{\text{T}}^{1})=-\frac{\sqrt{2}}{\pi}\int d\omega_1 d\omega_2k(\omega_1)k(\omega_2)k(\omega_1,\omega_2) \text{Im}[\frac{A_{\perp}A_{0}^*}{\mathcal{D}}-\frac{\bar{A}_{\perp}\bar{A}_{0}^*}{\bar{\mathcal{D}}}],\label{tpatrue1}\\
{\cal A}_{\text{T-true}}^{2}&=&\frac{1}{2}({\cal A}_{\text{T}}^{2}+\bar{{\cal A}}_{\text{T}}^{2})=-\frac{{2}}{\pi}\int d\omega_1 d\omega_2k(\omega_1)k(\omega_2)k(\omega_1,\omega_2) \text{Im}[\frac{A_{\perp}A_{||}^*}{\mathcal{D}}-\frac{\bar{A}_{\perp}\bar{A}_{||}^*}{\bar{\mathcal{D}}}],\label{tpatrue2}\\
{\cal A}_\text{T-fake}^{1}&=&\frac{1}{2}({\cal A}_{\text{T}}^{1}-\bar{{\cal A}}_{\text{T}}^{1})= -\frac{\sqrt{2}}{\pi}\int d\omega_1 d\omega_2k(\omega_1)k(\omega_2)k(\omega_1,\omega_2) \text{Im}[\frac{A_{\perp}A_{0}^*}{\mathcal{D}}+\frac{\bar{A}_{\perp}\bar{A}_{0}^*}{\bar{\mathcal{D}}}] \label{tpafake1},\\
{\cal A}_\text{T-fake}^{2}&=&\frac{1}{2}({\cal A}_{\text{T}}^{2}-\bar{{\cal A}}_{\text{T}}^{2})= -\frac{{2}}{\pi}\int d\omega_1 d\omega_2k(\omega_1)k(\omega_2)k(\omega_1,\omega_2) \text{Im}[\frac{A_{\perp}A_{||}^*}{\mathcal{D}}+\frac{\bar{A}_{\perp}\bar{A}_{||}^*}{\bar{\mathcal{D}}}] \label{tpafake2},
\end{eqnarray}
The subscripts ``true"  and ``fake"  refer to whether the asymmetry is due to a real $CP$ asymmetry or effects from
final-state interactions that are $CP$ symmetric.
The two asymmetries defined in Eqs.~(\ref{tpatrueave1}) and (\ref{tpatrue1}) are usually different from each other in the most $B_{(s)}$ meson decays,
as well as the two asymmetries in Eqs.~(\ref{tpatrueave2}) and (\ref{tpatrue2}).
They become equal when no direct $CP$ asymmetry occurs in the total decay rate, namely $\mathcal{D}=\bar{\mathcal{D}}$.

\section{Perturbative calculation}\label{sec:3}
For simplicity,
we will work in the rest frame of the $B$ meson.
In the light-cone coordinates,
the $B$ meson momentum $p_{B}$  can be parametrized as $p_{B}=\frac{m_{B}}{\sqrt 2}(1,1,\textbf{0}_{\rm T})$.
Considering  the four-body decay $B^0\to \rho^0\phi\to (\pi^+\pi^-)(K^+K^-)$ shown in Fig.~\ref{fig1},
we define the intermediate resonance $\rho(\phi)$ with the momentum $p(q)$, and the four final state mesons with the momentum $p_i(i=1,4)$,
satisfying the momentum conservation relations $p_{B}=p+q$, $p=p_1+p_2$,  $q=p_3+p_4$.
The momentum of the $\rho$ and $\phi$ can be written as
\begin{eqnarray}
p=\frac{m_{B}}{\sqrt2}(g^+,g^-,0_{\rm T}),\quad\quad
q=\frac{m_{B}}{\sqrt 2}(f^-,f^+,0_{\rm T}),
\end{eqnarray}
in which the factors $f^{\pm},g^{\pm}$ are related to the invariant masses of the meson pairs via $p^2=\omega_1^2$ and $q^2=\omega_2^2$,
\begin{eqnarray}\label{eq:epsilon}
g^{\pm}&=&\frac{1}{2}\left[1+\eta_1-\eta_2\pm\sqrt{(1+\eta_1-\eta_2)^2-4\eta_1}\right],\nonumber\\
f^{\pm}&=&\frac{1}{2}\left[1-\eta_1+\eta_2\pm\sqrt{(1+\eta_1-\eta_2)^2-4\eta_1}\right],
\end{eqnarray}
with the mass ratio $\eta_{1(2)}=\omega_{1(2)}^2/m^2_{B}$.
The corresponding  longitudinal polarization vectors of the $P$-wave $\pi\pi$ and $K\bar K$ pairs can be defined as
\begin{eqnarray}\label{eq:pq1}
\epsilon_{p}=\frac{1}{\sqrt{2\eta_1}}(g^+,-g^-,\textbf{0}_{T}),\quad
\epsilon_{q}=\frac{1}{\sqrt{2\eta_2}}(-f^-,f^+,\textbf{0}_{T}),
\end{eqnarray}
which obey the normalization $\epsilon_{p}^2=\epsilon_{q}^2=-1$  and the orthogonality
$\epsilon_{p}\cdot p=\epsilon_{q}\cdot q=0$.

The explicit expressions of the individual momenta $p_i$ can be derived from the
relations $p=p_1+p_2$ and $q=p_3+p_4$ together with the on-shell conditions
$p_i^2=m_i^2$ for the final state mesons,
\begin{eqnarray}\label{eq:p1p4}
p_1&=&\left(\frac{m_{B}}{\sqrt{2}}(\zeta_1+\frac{r_1-r_2}{2\eta_1})g^+,\frac{m_{B}}{\sqrt{2}}(1-\zeta_1+\frac{r_1-r_2}{2\eta_1})g^-,\textbf{p}_{T}\right),\nonumber\\
p_2&=&\left(\frac{m_{B}}{\sqrt{2}}(1-\zeta_1-\frac{r_1-r_2}{2\eta_1})g^+,\frac{m_{B}}{\sqrt{2}}(\zeta_1-\frac{r_1-r_2}{2\eta_1})g^-,-\textbf{p}_{T}\right),\nonumber\\
p_3&=&\left(\frac{m_{B}}{\sqrt{2}}(1-\zeta_2+\frac{r_3-r_4}{2\eta_2})f^-,\frac{m_{B}}{\sqrt{2}}(\zeta_2+\frac{r_3-r_4}{2\eta_2})f^+,\textbf{q}_{T}\right),\nonumber\\
p_4&=&\left(\frac{m_{B}}{\sqrt{2}}(\zeta_2-\frac{r_3-r_4}{2\eta_2})f^-,\frac{m_{B}}{\sqrt{2}}(1-\zeta_2-\frac{r_3-r_4}{2\eta_2})f^+,-\textbf{q}_{T}\right),\nonumber\\
\textbf{p}_{\rm T}^2&=&\zeta_1(1-\zeta_1)\omega_1^2-\frac{r_1+r_2}{2\eta_1}+\frac{(r_1-r_2)^2}{4\eta_1^2},\nonumber\\
\textbf{q}_{\rm T}^2&=&\zeta_2(1-\zeta_2)\omega_2^2-\frac{r_3+r_4}{2\eta_2}+\frac{(r_3-r_4)^2}{4\eta_2^2},
\end{eqnarray}
with the mass ratios $r_i=m_i^2/m^2_{B}$, $m_i$ being the masses of the final state mesons,
and the term $\zeta_1+\frac{r_1-r_2}{2\eta_1}=p_1^+/p^+$ $(\zeta_2+\frac{r_3-r_4}{2\eta_2}=p_3^-/q^-)$ characterizing
the momentum fraction for one of pion-pion (kaon-kaon) pair.

It is easy to obtain the relation between the meson momentum fractions $\zeta_{1,2}$
and the polar angle $\theta_{1,2}$ in the dimeson rest frame in Fig.~\ref{fig1},
\begin{eqnarray}\label{eq:cos}
2\zeta_{1}-1=\sqrt{1+4\alpha_1}\cos\theta_1, ~\quad
2\zeta_{2}-1=\sqrt{1+4\alpha_2}\cos\theta_2,
\end{eqnarray}
with the two factors
\begin{eqnarray}\label{eq:alpha12}
\alpha_1=-\frac{r_1+r_2}{2\eta_1}+\frac{(r_1-r_2)^2}{4\eta_1^2},\quad
\alpha_2=-\frac{r_3+r_4}{2\eta_2}+\frac{(r_3-r_4)^2}{4\eta_2^2},
\end{eqnarray}
and the bound
\begin{eqnarray}
\zeta_{1\text{max,min}}=\frac{1}{2}\left[1\pm\sqrt{1+4\alpha_1}\right],~\quad
\zeta_{2\text{max,min}}=\frac{1}{2}\left[1\pm\sqrt{1+4\alpha_2}\right].
\end{eqnarray}

As illustrated in Fig.~\ref{fig2},
there are eight types of Feynman diagrams contributing to the hard kernels $H$ of
 the four-body decays $B_{(s)}\to R_1R_2 \to (\pi\pi)(KK)$ at leading order in the PQCD approach,
which can be classified into three types:
the factorizable emission diagrams (Figs.~\ref{fig2}(a) and~\ref{fig2}(b));
the nonfactorizable emission diagrams (Figs.~\ref{fig2}(c) and~\ref{fig2}(d));
and the annihilation diagrams (Figs.~\ref{fig2}(e)-\ref{fig2}(h)).
For the evaluation of the $H$,
we also need to define three valence quark momenta labelled by $k_i$ $(i=B,p,q)$ in each meson as
\begin{eqnarray}
k_{B}&=&\left(0,x_B p_B^+ ,\textbf{k}_{B \rm T}\right),\quad
k_p= \left( x_1 p^+,0,\textbf{k}_{1{\rm T}}\right),\quad\quad
k_q=\left(0,x_2q^-,\textbf{k}_{2{\rm T}}\right),\label{eq:mom-B-k}
\end{eqnarray}
with the parton momentum fraction $x_i$, and the parton transverse momentum $\textbf{k}_{\rm iT}$.
The small components $k^-_{p}$ and  $k^+_{q}$ in Eq.~(\ref{eq:mom-B-k})
have been dropped in our calculation
because $k_p$ and $k_q$ are aligned with the meson pairs in the plus and minus direction.
We also neglect the contribution from the $k^+_B$ component since it
does not appear in the hard kernels for dominant factorizable contributions.

In order to calculate the different helicity amplitudes,
we first give the weak effective Hamiltonian ${\cal H}_{eff}$ of the considered four-body decays induced by the $b\to q$ $(q=s,d)$ transition,
\begin{eqnarray}\label{eq:heff}
{\cal H}_{eff} &=& \frac{G_{F}}{\sqrt{2}}     \Bigg\{ V_{ub} V_{uq}^{\ast} \Big[
     C_{1}({\mu}) O^{u}_{1}({\mu})  +  C_{2}({\mu}) O^{u}_{2}({\mu})\Big]
  -V_{tb} V_{tq}^{\ast} \Big[{\sum\limits_{i=3}^{10}} C_{i}({\mu}) O_{i}({\mu})
  \Big ] \Bigg\} + \mbox{H.c.} ,
 \label{eq:heff}
\end{eqnarray}
with the Fermi constant $G_{F}=1.166 39\times 10^{-5}$ GeV$^{-2}$,
Wilson coefficients $C_i(\mu)$ at the renormalization scale $\mu$,
the local four-quark operators $O_{i}$ ($i=1,...,10$)~\cite{rmp68-1125}
and the CKM matrix elements $V_{ij}$.

According to the above Eq.~(\ref{eq:heff}),
each considered decay channel may receive contributions from one or more terms proportional to different Wilson coefficients $C_i$.
The total decay amplitudes of the $B_{(s)} \to \rho\phi\to(\pi\pi)(K {\bar K})$ at LO in the PQCD approach can then be written as
\begin{eqnarray}
\label{eq:b+r0phi}
A_h(B^+ \to \rho^+\phi\to(\pi^+\pi^-)(K^+K^-)&=&
 -\frac{G_F}{\sqrt{2}}V_{tb}^*V_{td}\Big [ \left(C_3+\frac{C_4}{3}+C_5+\frac{C_6}{3}-\frac{C_7}{2}-\frac{C_{8}}{6}- \frac{C_9}{2}-\frac{C_{10}}{6} \right )F^{LL,h}_{e\rho}  \non
&+&\left(C_4- \frac{C_{10}}{2}\right )M^{LL,h}_{e\rho}+\left(C_6- \frac{C_{8}}{2}\right )M^{SP,h}_{e\rho}\Big ]\Big\},
\end{eqnarray}
\begin{eqnarray}
A_h(B^0 \to \rho^0\phi\to(\pi^+\pi^-)(K^+K^-)&=&
 -\frac{G_F}{2}V_{tb}^*V_{td}\Big [ \left(-C_3-\frac{C_4}{3}-C_5-\frac{C_6}{3}+\frac{C_7}{2}+\frac{C_{8}}{6}+ \frac{C_9}{2}+\frac{C_{10}}{6} \right )F^{LL,h}_{e\rho}  \non
&-&\left(C_4- \frac{C_{10}}{2}\right )M^{LL,h}_{e\rho}-\left(C_6- \frac{C_{8}}{2}\right )M^{SP,h}_{e\rho}\Big ]\Big\},
\end{eqnarray}
\begin{eqnarray}
A_h(B_s^0 \to \rho^0\phi\to (\pi^+\pi^-)(K^+K^-))&=&\frac{G_F} {2}\Big\{V_{ub}^*V_{us}\Big[\left(C_1+\frac{C_2}{3}\right )F^{LL,h}_{e\phi}
+C_2M^{LL,h}_{e\phi}\Big ]  \non
 &-&V_{tb}^*V_{ts}\Big [  \frac{3}{2}\left(C_7+\frac{C_8}{3}+C_9+\frac{C_{10}}{3} \right )F^{LL,h}_{e\phi}
 +\frac{3C_{10}}{2}M^{LL,h}_{e\phi}+\frac{3C_{8}}{2}M^{SP,h}_{e\phi}\Big ]\Big\},\label{eq:b0r0phi}\non
\end{eqnarray}
with $h=0,\|,\bot$.
The individual decay amplitude appeared in the above equations, such as $F^{LL,h}_{e\rho}$ and $M^{LL,h}_{e\rho}$, $M^{SP,h}_{e\rho}$,
is obtained by evaluating the Feynman diagrams in Fig.~\ref{fig2} analytically.
The term $F^{LL,h}_{e\rho}$ ( $M^{SP,h}_{e\rho}$), for example,
represents the contribution from the factorizable (nonfactorizable) emission diagrams with $(V-A)\otimes(V-A)$ $((S-P)\otimes(S+P))$ current.
The explicit expressions of $F^{LL,h}_{e\rho}$ and other decay amplitudes  can be found easily in Ref.~\cite{zjhep}.
The helicity amplitudes of the $S$-wave decays have been collected in APPENDIX~\ref{samp}.

As shown in Eq.~(\ref{amplitude}),
the DAs of the initial $B$ meson and the final state meson pairs are the most important nonperturbative inputs in the PQCD approach.
For the $B$ meson,
we adopt the form widely used in the literature~\cite{Li:2003yj,Kurimoto:2001zj},
and more alternative models of the $B$ meson DA and the subleading contributions can be found in Refs.~\cite{prd102011502,prd70074030,Li:2012md,Li:2014xda,Li:2012nk,2207.02691,jhep05-157,npb990-116175}.
The $S$- and $P$-wave two-pion (kaon) DAs, as well as the related  time-like form factors are summarized in APPENDIX~\ref{TMDAs}.

\section{Numerical Analysis}\label{sec:4}
In this section,
we work out a number of physical observables for the $B_{(s)}\to (\pi\pi)(K\bar K)$ decays,
such as the branching ratios, polarization fractions, direct $CP$ asymmetries, together with TPAs.
We firstly show the input parameters adopted in our numerical analysis in Table~\ref{tab:constant},
including the decay constants~\cite{prd76-074018,prd104-096014},
the meson masses, the decay widths, the lifetimes and  Wolfenstein parameters~\cite{pdg2020}.

The differential branching fraction of the $B_{(s)} \rightarrow (\pi\pi)(K{\bar K})$ in the $B_{(s)} $ meson rest frame
can be expressed as:
\begin{eqnarray}\label{eq:decayrate}
\frac{d^5\mathcal{B}}{d\theta_1 d\theta_2 d\varphi d\omega_1d\omega_2}
=\frac{\tau_{B_{(s)}} k(\omega_1)k(\omega_2)k(\omega_1,\omega_2)}{16(2\pi)^6m_{B_{(s)}}^2} |A|^2, 
\end{eqnarray}
with the $B_{(s)}$ meson lifetime $\tau_{B_{(s)}}$.
It has been confirmed that Eq.~(\ref{eq:decayrate}) is equivalent to those in Refs.~\cite{prd85-094019,plb770-348}
by appropriate variable changes.
Replacing the helicity angle $\theta_{1(2)}$ by the meson momentum fraction $\zeta_{1(2)}$ via Eq.~(\ref{eq:cos}),
the Eq.~(\ref{eq:decayrate}) is turned into
\begin{eqnarray}\label{eq:decayrate1}
\frac{d^5\mathcal{B}}{d\zeta_1d\zeta_2d \omega_1d \omega_2d\varphi}=
\frac{\tau_{B_{(s)}} k(\omega_1)k(\omega_2)k(\omega_1,\omega_2)}{4(2\pi)^6m_{B_{(s)}}^2\sqrt{1+4\alpha_1}\sqrt{1+4\alpha_2}}|A|^2,
\end{eqnarray}
where the total decay amplitude $A$ can be written as a coherent sum of the $P$-, $S$-, and double $S$-wave components
with $\zeta_{1(2)}$ and $\varphi$ dependencies
\begin{eqnarray}\label{eq:allampli}
A&=&A_0\frac{2\zeta_1-1}{\sqrt{1+4\alpha_1}}\frac{2\zeta_2-1}{\sqrt{1+4\alpha_2}}
+A_{\parallel}2\sqrt{2}\sqrt{\frac{\zeta_1(1-\zeta_1)+\alpha_1}{1+4\alpha_1}}
\sqrt{\frac{\zeta_2(1-\zeta_2)+\alpha_2}{1+4\alpha_2}}\cos\varphi \nonumber\\
&&
+i A_{\perp}2\sqrt{2}\sqrt{\frac{\zeta_1(1-\zeta_1)+\alpha_1}{1+4\alpha_1}}
\sqrt{\frac{\zeta_2(1-\zeta_2)+\alpha_2}{1+4\alpha_2}}\sin\varphi\non
&& +A_{VS}\frac{2\zeta_1-1}{\sqrt{1+4\alpha_1}}
+A_{SV}\frac{2\zeta_2-1}{\sqrt{1+4\alpha_2}}+A_{SS}.
\end{eqnarray}

We can obtain the branching ratio form according to the Eq.~(\ref{eq:decayrate1}),
\begin{eqnarray}\label{eq:brsss}
\mathcal{B}_h=\frac{\tau_{B_{(s)}}}{4(2\pi)^6m_{B_{(s)}}^2}\frac{2\pi}{9}C_h
\int d\omega_1d\omega_2k(\omega_1)k(\omega_2)k(\omega_1,\omega_2)|A_h|^2.
\end{eqnarray}
The coefficients $C_h$ are the results of the integrations over $\zeta_{1},\zeta_{2},\varphi$ in terms of Eq.~(\ref{eq:brsss}) and listed as follows,
\begin{eqnarray}\label{eq:radii}
C_h=\left\{\begin{aligned}
&(1+4\alpha_1)(1+4\alpha_2), \quad\quad\quad  &h=0,\parallel,\perp, \\
&3(1+4\alpha_{1,2}) ,\quad\quad\quad  &h=VS,SV, \\
&9,   &h=SS. \\
\end{aligned}\right.
\end{eqnarray}

The $CP$-averaged branching ratio, the direct $CP$ asymmetries in each component and the overall  asymmetry can then be defined as below,
\begin{eqnarray}
\mathcal{B}_h^{\rm avg}=\frac{1}{2}(\mathcal{B}_h+\mathcal{\bar{B}}_h),
\quad \mathcal{A}^{\text{dir}}_h=\frac{\mathcal{\bar{B}}_h-\mathcal{B}_h}{\mathcal{\bar{B}}_h+\mathcal{B}_h},\quad
\mathcal{A}^{\text{dir}}_{CP}=\frac{\sum_h \mathcal{\bar{B}}_h
-\sum_h \mathcal{B}_h}{\sum_h \mathcal{\bar{B}}_h+\sum_h \mathcal{B}_h},
\end{eqnarray}
where $\mathcal{\bar{B}}_h$ is the branching ratio of the corresponding $CP$-conjugate channel.

For the $B\to VV$ decays, the additional polarization fractions $f_{\lambda}$ with $\lambda=0$, $\parallel$,
and $\perp$, are described as
\begin{eqnarray}\label{pol}
f_{\lambda}=\frac{|A_{\lambda}|^2}{|A_0|^2
+|A_{\parallel}|^2+|A_{\perp}|^2},
\end{eqnarray}
with the normalisation relation $f_0+f_{\parallel}+f_{\perp}=1$.

\begin{table}
\caption{The decay constants are taken from Refs.~\cite{prd76-074018,prd104-096014}.
Other parameters are from PDG 2022~\cite{pdg2020}. }
\label{tab:constant}
\centering
\begin{tabular*}{14.5cm}{@{\extracolsep{\fill}}lllll}
  \hline\hline
\text{Wolfenstein parameters}
& $\lambda=0.22650$  & $A=0.790$  &$\bar{\rho}=0.141$ & $\bar{\eta}=0.357$ \\[1ex]
\text{Mass (\text{GeV})}
&$m_{B}=5.28$  &$m_{B_s}=5.37$  & $m_{\pi^{\pm}}=0.140$  &$m_{K^{\pm}}=0.494$ \\[1ex]
& $m_{\pi^0}=0.135$ & $m_{K^0}=0.498$\\[1ex]
\text{Decay constants (\text{GeV})}
& $f_{B}=0.21$& $f_{B_s}=0.23$   &$f_{\phi}=0.215$ &$f_{\phi}^T=0.186$ \\[1ex]
&$f_{\rho}=0.216$&$f_{\rho}^T=0.184$\\
\text{Decay width (\text{MeV})}
& $\Gamma_{\phi}=4.25$ & $\Gamma_{\rho}=149.1$\\[1ex]
\text{Lifetime (ps)}
& $\tau_{B^0}=1.519$ & $\tau_{B^{\pm}}=1.638$ & $\tau_{B_s}=1.51$ \\[1ex]
\hline\hline
\end{tabular*}
\end{table}

\subsection{$S$-wave contributions}\label{s-wave}
\begin{table}[!htbh]
\caption{PQCD predictions for the  branching ratios of  the $B_{(s)}\rightarrow [VS,SV,SS]\to (\pi\pi)(K \bar K)$ decays, with $S=f_0$ and $V=\rho,\phi$.
The theoretical uncertainties are attributed to the variations of the shape parameter $\omega_{B_{(s)}}$
in the $B_{(s)}$ meson DA, of the Gegenbauer moments in various twist DAs of $K {\bar K}$ and $\pi\pi$ pairs,
and of the hard scale $t$ and the QCD scale $\Lambda_{\rm QCD}$.}
\label{tab:brfour}
\begin{tabular*}{12cm}{@{\extracolsep{\fill}}ll} \hline\hline
decay modes               &{\rm PQCD predictions}                                       \\ \hline
$B^+\to \rho^+f_0\to (\pi^+\pi^0)(K^+K^-)$          &$(1.25_{-0.45-0.20-0.25}^{+0.59+0.22+0.15})\times10^{-7}$                     \\
$B^0\to \rho^0 f_0\to (\pi^+\pi^-)(K^+K^-)$          &$(3.01_{-0.39-0.95-0.18}^{+0.50+1.34+0.49})\times10^{-9}$                     \\
$B^0_s\to \rho^0 f_0\to (\pi^+\pi^-)(K^+K^-)$          &$(4.01_{-0.76-0.35-0.34}^{+1.08+0.38+0.35})\times10^{-9}$                     \\
$B^0\to f_0 \phi\to   (\pi^+\pi^-)(K^+K^-)$          &$(0.47_{-0.13-0.24-0.06}^{+0.19+0.20+0.12})\times10^{-9}$                     \\
$B^0_s \to f_0 \phi\to   (\pi^+\pi^-)(K^+K^-)$          &$(4.09_{-1.58-1.02-1.67}^{+2.24+1.15+1.08})\times10^{-7}$                     \\
$B^0\to f_0 f_0 \to   (\pi^+\pi^-)(K^+K^-)$          &$(1.93_{-0.67-0.33-0.23}^{+1.05+0.66+0.44})\times10^{-10}$                     \\
$B^0_s\to f_0 f_0 \to   (\pi^+\pi^-)(K^+K^-)$          &$(7.19_{-2.28-0.60-2.63}^{+4.32+0.71+3.98})\times10^{-8}$                     \\
\hline\hline
\end{tabular*}
\end{table}

The PQCD predictions for the $CP$-averaged branching ratios of the $S$-wave decays are summarized in Table~\ref{tab:brfour},
in which the theoretical uncertainties are derived  from three different sources.
The first error results from the parameter of the wave function of the initial state $B_{(s)}$ meson,
$\omega_B=0.40\pm 0.04$ and $\omega_{B_s}=0.48\pm 0.048$~\cite{2012-15074,plb504-6,prd63-074009}.
The second one comes from the Gegenbauer moments in the two-meson DAs given in Eq.~(\ref{eq:gen}),
and the last one is caused by the variation of the hard scale $t$ from $0.75t$ to $1.25t$ (without changing $1/b_i$)
and the QCD scale $\Lambda_{\rm QCD}=0.25\pm0.05$~GeV, which characterizes the effect of the next-to-leading-order QCD contributions.
The three uncertainties are comparable, and their combined impacts could exceed $50\%$,
implying that the nonperturbative parameters in the DAs of the initial and final states need to be constrained more precisely,
and the higher-order corrections to four-body $B$ meson decays are critical.

Although the quark model has achieved great successes,
the identification of scalar mesons is a long-standing puzzle,
and the underlying structure of scalar mesons is not well established on the theoretical side
(for a review, see Ref.~\cite{pdg2020}).
At present, two main scenarios have been proposed to classify the light scalar resonances~\cite{prd73-014017}.
The Scenario-I (S-I) is based on the naive two-quark model,
and the light scalar mesons below or near 1 GeV like $f_0(980)$ are regarded as the lowest lying states.
While in Scenario-II (S-II), the $f_0(980)$ is identified as the predominant four-quark state $q^2{\bar q}^2$, and the scalars above 1 GeV are treated as the ground $q {\bar q}$ states.
Since it is difficult for us to study these $S$-wave decays based on the four-quark picture,
we shall consider the conventional $q{\bar q}$ assignment for the light scalar $f_0(980)$ to give several quantitative predictions.
In Scenario-I,
$f_0(980)$ is mainly treated as an $s{\bar s}$ state,
which has been supported by $D_s^+\to f_0\pi$, $\phi\to f_0\gamma$ decays~\cite{pdg2020}.
However, there also exists some experimental evidences indicating that $f_0(980)$ is not purely an $s{\bar s}$ state.
For example,
the observation of $\Gamma(J/\psi \to f_0 \omega) \simeq \frac{1}{2}\Gamma(J/\psi \to f_0 \phi)$~\cite{pdg2020}
clearly shows the existence of the non-strange and strange quark content.
Therefore,
 $f_0(980)$ should be a mixture of $n {\bar n}=\frac{1}{\sqrt{2}}(u{\bar u}+d{\bar d})$ and $s{\bar s}$,
\begin{eqnarray}\label{f0mix}
|f_0\rangle=|n {\bar n}\rangle {\rm sin} \theta+|s {\bar s}\rangle{\rm cos}\theta.
\end{eqnarray}
The value of the mixing angle $\theta$ in the above equation has not been determined precisely so far,
which is suggested to be in the wide ranges of $25^{\circ} <\theta < 40^{\circ}$ and $140^{\circ} <\theta < 165^{\circ}$~\cite{prd67-034024,pan65-497,plb609-291}.
For simplicity, we will adopt the value $\theta=145^{\circ}$~\cite{epjc82-59,prd103-113005} in our calculation.
In our previous works~\cite{zjhep,Li:2021qiw,prd105-053002,prd105-093001},
the scalar meson $f_0(980)$ is usually considered as the pure $s {\bar s}$ state.
Furthermore,
there also exist some theoretical studies on the $B_{(s)}$ meson decays involving $f_0(980)$ in the final states
based on the assumption that $f_0(980)$ is a pure $s {\bar s}$ density operator~\cite{prd81-074001,prd83-094027,epjc80-554}.
In Ref.~\cite{epjc83-974},
on the other hand,
we found that the contribution from the $f_0=(u {\bar u}+d {\bar d})/\sqrt{2}$ component is significant
in the decays like $B^0\to \rho^0f_0$ governed by  $B \to f_n$ transition form factor.
Thus,
the mixing effect shown in Eq.~(\ref{f0mix}) should also be taken into account in this work.

The two-body branching ratio ${\cal B}(B_{(s)} \to R_1R_2)$ can usually be extracted from the corresponding four-body decay modes in Table~\ref{tab:brfour}
under the narrow width approximation:
\beq
\label{2body}
\mathcal{B}(B_{(s)} \rightarrow R_1R_2\rightarrow (\pi\pi)( K\bar K))&\approx& \mathcal{B}(B_{(s)} \rightarrow R_1R_2)\times
\mathcal{B}(R_1 \rightarrow \pi\pi)\times \mathcal{B}(R_2 \rightarrow K\bar K).
\eeq

The $\cal B$ of the three-body decay $B^0 \to \rho^0(f_0\to)\pi^+\pi^-$ can then be calculated as follows:
\begin{eqnarray}
\label{f03}
{\cal B}(B^0 \to \rho^0(f_0\to)\pi^+\pi^-)=\frac{{\cal B}(B^0\to \rho^0f_0\to(\pi^+\pi^-)(K^+K^-))}{{\cal B}(\rho^0\to \pi^+\pi^-)}\cdot R_{\pi/K}=(0.09^{+0.05}_{-0.03})\times 10^{-7},
\end{eqnarray}
with the ratio $R_{\pi/K}=\frac{{\cal B}(f_0\to \pi^+\pi^-)}{{\cal B}(f_0\to K^+K^-)}$.
Recent years,
$BABAR$~\cite{prd74-032003} and BES~\cite{prd70-092002,prd72-092002} Collaborations have performed systematically measurements on the ratio of the
partial decay width of $f_0\to K^+K^-$ to $f_0\to \pi^+\pi^-$,
\begin{eqnarray}\label{rkpiexp}
R^{\rm exp}_{K/\pi}=
\left\{\begin{array}{ll}
0.69\pm 0.32      & {BABAR},\\
0.25^{+0.17}_{-0.11}     & {\rm BES},\\
\end{array} \right.
\end{eqnarray}
and we have adopted their average value $R^{\rm exp}_{K/\pi}=0.35\pm 0.11$~\cite{prd92-032002} in Eq.~(\ref{f03}).
The $B^0 \to \rho^0 f_0$ decay mode has also been studied in the two-body framework
within the PQCD~\cite{prd82-034036} and QCDF~\cite{prd87-114001} approaches.
In the narrow-width limit,
one can get the following branching fractions
\begin{eqnarray}
{\cal B}(B^0 \to \rho^0(f_0\to)\pi^+\pi^-)=
\left\{\begin{array}{ll}
(0.10^{+0.15}_{-0.00})\times 10^{-7}     & {\rm QCDF} ,\\
(1.65^{+1.00}_{-0.84})\times 10^{-7}      & {\rm PQCD} ,\\
\end{array} \right.
\end{eqnarray}
where ${\cal B}(f_0\to\pi^+\pi^-)=0.5$ \cite{prd87-114001} is used.
Our calculation ${\cal B}=(0.09^{+0.05}_{-0.03})\times 10^{-7}$ is consistent well with the  QCDF prediction~\cite{prd87-114001},
but far from the previous PQCD value~\cite{prd82-034036}.
Nonetheless,
all these theoretical predictions are much smaller than the current experimental data ${\cal B}^{\rm exp}=(7.8\pm 2.5)\times 10^{-7}$~\cite{pdg2020},
which may be clarified in the following form.
First,
the $B^0 \to \rho^0 (f_0\to)\pi^+\pi^-$ decay is ascribed to the involved color suppressed tree contributions.
Since only leading order diagrams have been concerned in the current work,
it indicates that this decay might receive substantial  next-to-leading-order (NLO) corrections.
Besides,
as shown in Ref.~\cite{epjc83-974},
the calculated $\cal B$ of the decay $B^0\to f_0\rho^0$
is sensitive to the Gegenbauer moment $a_S$ and the mixing angel $\theta$.
It is expected that maybe
we can fit the related non-perturbative parameters with abundant data to match the experiment
when NLO corrections to four-body decays in the PQCD framework are considered,
which goes beyond the scope of the present work and should be left for the future studies.

We remark that the $K {\bar K}$  invariant mass of the $S$-wave decays has been limited
in a narrow window of $\pm 30$ MeV around the known $\phi$ mass in this study.
As claimed in Refs.~\cite{prd105-053002,prd105-093001},
the contribution of scalar resonance $f_0(980)$ relies on the final-state invariant mass range strongly, since it has a wide decay width.
We have recalculated the $\cal B$ of the decay $B^0\to \rho^0f_0\to(\pi^+\pi^-)(K^+K^-)$
by enlarging the $S$-wave $K {\bar K}$ invariant mass range from $[m_\phi-30 {\rm MeV}, m_\phi+30{\rm MeV}]$ to $[2m_K, m_B-m_\rho]$:
${\cal B}=2.95\times 10^{-8}$.
The corresponding branching fraction of the three-body decay $B^0 \to \rho^0(f_0\to)\pi^+\pi^-$ is then estimated to be $0.84\times 10^{-7}$,
which is larger than the value in Eq.~(\ref{f03}) by almost one order.
It should also be noted that,
strictly speaking,
 the narrow width approximation has not been fully justified since such approximation has its scope of application.
As claimed in Refs.~\cite{Cheng:2020iwk,Cheng:2020mna}, for the broad scalar intermediate states like $f_0(980)$,
the narrow-width approximation should be corrected by including the finite-width effects.
The result ${\cal B}=(0.09^{+0.05}_{-0.03})\times 10^{-7}$ evaluated from the ${\cal B}(B^0\to \rho^0f_0\to(\pi^+\pi^-)(K^+K^-))$
 may suffer from a large uncertainty due to the finite-width effects of the scalar resonance.
Therefore,
we hope that the future experiments can perform a direct measurement on the four-body decay $B^0\to \rho^0f_0\to(\pi^+\pi^-)(K^+K^-)$.

Relying on the fraction ${\cal{B}}(\phi\to K^+K^-)=49.2\%$~\cite{pdg2020},
we can  extract  the ${\cal B}(B_s^0 \to \phi(f_0\to)\pi^+\pi^-)$
from the four-body decay $B^0_{s}\to f_0\phi\to (\pi^+\pi^-)( K^+K^-)$ in Table~\ref{tab:brfour} under the narrow width limit:
\begin{eqnarray}\label{bsf0phi}
{\cal B}(B_s^0 \to \phi (f_0\to)\pi^+\pi^-)=\frac{{\cal B}(B^0_{s}\to f_0 \phi \to (\pi^+\pi^-) ( K^+K^-))}{{\cal{B}}(\phi\to K^+K^-)}=(0.83^{+0.54}_{-0.51})\times 10^{-6}.
\end{eqnarray}
Although the above theoretical cental value is a bit smaller than the experimental data ${\cal B}^{\rm exp}=(1.12\pm 0.21)\times 10^{-6}$~\cite{pdg2020},
our result can still accommodate the current measurement with large uncertainties,
and also comparable with the previous three-body PQCD result~\cite{epjc81-91}.
In Ref.~\cite{epjc82-59},
the authors have studied the branching fraction of the two-body decay $B_s^0 \to \phi f_0$ in PQCD approach,
and one can obtain ${\cal B}(B_s^0 \to \phi (f_0\to)\pi^+\pi^-)=(0.24^{+0.21}_{-0.14})\times 10^{-6}$ according to Eq.~(\ref{2body}).
It is shown that our calculation ${\cal B}=(0.83^{+0.54}_{-0.51})\times 10^{-6}$ presented in Eq.~(\ref{bsf0phi})
is a bit larger than the converted value $(0.24^{+0.21}_{-0.14})\times 10^{-6}$~\cite{epjc82-59} from the previous two-body PQCD result by a factor of $\sim 3$ ,
but more close to the experimental data.
As already stressed previously that
 in fact the narrow width approximation is not exactly valid for the broad intermediate states like $f_0(980)$.
For these resonances,
the finite-width effects is significant and should be considered.
Thus,
the above comparisons is just a rough estimate for a cross-checking.
Overall,
since the property of the scalar resonance $f_0(980)$ is not well understood,
and  both the theoretical and experimental uncertainties are relatively large,
all the above issues need to be further clarified in the future.

\subsection{Branching ratios and polarization fractions of two-body $B_{(s)} \to \rho\phi$ decays}
On basis of the narrow width approximation Eq.(\ref{2body}),
the branching ratios of the two-body decays $B_{(s)} \to \rho\phi$ have been extracted in Table~\ref{tab:brtwo}.
The polarization fractions $f_i$ ($i=0,\|,\bot$) of the two-body  $B_{(s)} \to \rho\phi$ decays
calculated in this work have also been listed in Table~\ref{tab:brtwo}.
For a comparison, the updated predictions in the QCDF~\cite{prd80-114026}, the previous predictions in the PQCD approach~\cite{prd91-054033}, SCET~\cite{prd96-073004} and FAT~\cite{epjc77-333} are also displayed in Table~\ref{tab:brtwo},
and the experimental results for branching ratios are taken from PDG 2022~\cite{pdg2020}.

\begin{table}[!htbh]
\caption{Branching ratios and polarization fractions  of the two-body $B_{(s)}\rightarrow \rho \phi$ decays.
For a comparison, we also list the results from the previous PQCD \cite{prd91-054033},
QCDF \cite{prd80-114026}, SCET \cite{prd96-073004}, and FAT \cite{epjc77-333}.
The world averages of experimental data are taken from PDG 2022~\cite{pdg2020}.
The sources of the theoretical errors are the same as in Table~\ref{tab:brfour}.}
\label{tab:brtwo}
\begin{ruledtabular}  \begin{threeparttable}
\setlength{\tabcolsep}{1mm}{ \begin{tabular}[t]{lcccc }
Modes     & $\mathcal{B}(10^{-6})$       & $f_0(\%)$  & $f_\|(\%)$  & $f_\perp(\%)$
                   \\ \hline
$B_s^0\rightarrow\rho^0 \phi$    &$0.28_{-0.07-0.01-0.02}^{+0.11+0.02+0.03}$       &$86.58_{-0.13-0.52-0.19}^{+0.25+0.56+0.46}$
                  &$6.45_{-0.11-0.26-0.17}^{+0.05+0.27+0.09}$     &$6.98_{-0.14-0.30-0.24}^{+0.07+0.30+0.11}$ \\
PQCD~\cite{prd91-054033}      &$0.23^{+0.15}_{-0.05}$         &$86 \pm 1$      & $\cdots$  &$8.89_{-1.06}^{+0.80}$
                                          \\
QCDF~\cite{prd80-114026}        & $0.18^{+0.09}_{-0.04}$           & $88^{+2}_{-18}$           & $\cdots$  &$\cdots$             \\
SCET~\cite{prd96-073004}        & $0.36\pm0.05$                     & $100$             & $\cdots$  &$\cdots$       \\
FAT~\cite{epjc77-333}           &$0.07\pm0.03$                     & $\cdots$              & $\cdots$  &$\cdots$           \\
Data~\cite{pdg2020}                           & $0.27\pm0.08$                     & $\cdots$              & $\cdots$  &$\cdots$       \\\hline
$B^0\rightarrow\rho^0 \phi$    &$0.006_{-0.001-0.002-0.001}^{+0.001+0.002+0.001}$       &$85.47_{-0.46-6.52-6.46}^{+0.00+3.95+4.26}$
&$7.73_{-0.06-2.10-3.08}^{+0.25+3.46+4.36}$     &$6.80_{-0.00-1.86-1.19}^{+0.21+3.04+2.06}$ \\
PQCD~\cite{prd91-054033}      &$0.013^{+0.007}_{-0.006}$         &$95\pm 1$      & $\cdots$    &$2.36_{-0.76}^{+1.08}$
                                          \\
SCET~\cite{prd96-073004}        & $\approx 0.002$                     & $100$             & $\cdots$  &$\cdots$       \\
FAT~\cite{epjc77-333}           &$0.03\pm0.01$                     & $\cdots$              & $\cdots$  &$\cdots$           \\
Data~\cite{pdg2020}                           & $<0.33$                     & $\cdots$              & $\cdots$  &$\cdots$       \\\hline
$B^+\rightarrow\rho^+ \phi$    &$0.013_{-0.002-0.004-0.002}^{+0.002+0.004+0.002}$       &$85.47_{-0.46-6.52-6.46}^{+0.00+3.95+4.26}$
&$7.73_{-0.06-2.10-3.08}^{+0.25+3.46+4.36}$     &$6.80_{-0.00-1.86-1.19}^{+0.21+3.04+2.06}$ \\
PQCD~\cite{prd91-054033}      &$0.028^{+0.015}_{-0.013}$         &$95^{+1}_{-2}$      & $\cdots$ &$2.36_{-0.76}^{+1.08}$
                                          \\
SCET~\cite{prd96-073004}        & $0.005\pm0.001$                     & $100$             & $\cdots$  &$\cdots$       \\
FAT~\cite{epjc77-333}           &$0.06\pm0.02$                     & $\cdots$              & $\cdots$  &$\cdots$           \\
Data~\cite{pdg2020}                           & $<3.0$                     & $\cdots$              & $\cdots$  &$\cdots$   \\
\end{tabular}}
\end{threeparttable}
\end{ruledtabular}
\end{table}

\begin{table}[htbp!]
	\centering
	\caption{Branching ratios of the four-body decay $B_s^0\to\rho^0 \phi\to (\pi^+\pi^-)(K^+K^-)$ from different topology diagrams.
   FE and NFE represent the contributions from factorizable emission and nonfactorizable emission diagrams,
respectively.}
\begin{ruledtabular}
\begin{threeparttable}
    \setlength{\tabcolsep}{1mm}{
	\begin{tabular}{lccccccc}
\multirow{2}{*}{channel}                &\multicolumn{3}{c}{Tree ${\cal B}(10^{-6})$}                       &\multicolumn{3}{c}{Penguin ${\cal B}(10^{-6})$}\cr\cline{2-7}
	                                     &${\rm FE}$    &${\rm NFE}$   &{\rm Total}   &${\rm FE}$     &${\rm NFE}$  &{\rm Total}\cr \hline
$B_s^0\to\rho^0 \phi\to (\pi^+\pi^-)(K^+K^-)$       &$0.013$      &0.018     & $0.031$      &$0.214$   &0.001 & $0.215$    \\
\end{tabular}}
\end{threeparttable}
\end{ruledtabular}
\label{rhophitp}
\end{table}
Most of the theoretical predictions of ${\cal B}(B_s^0\to \rho^0 \phi)$ agree well with the current data within errors.
The calculated branching ratio $(0.28^{+0.12}_{-0.07})\times 10^{-6}$ of the decay $B_s^0\to \rho^0 \phi$ is much smaller
than those of other $b\to s$ transition processes, such as $B^0_s\to K^*K^*$ decays.
To see clearly the contributions from different topology diagrams,
we show the explicit numerical results of the  $B_s^0\to \rho^0 \phi$ decay  in Table.~\ref{rhophitp},
in which we just quote the central values.
We know that $|V^*_{tb}V_{ts}|$ and $|V^*_{ub}V_{us}|$ are ${\cal O}(\lambda^2)$ and ${\cal O}(\lambda^4)$ respectively, with $\lambda\sim 0.22$.
This implies that the tree operators of the $b\to s$ transition decays like $B_s^0\to \rho^0 \phi$ are highly suppressed by the CKM matrix elements $|V^*_{ub}V_{us}|$.
Furthermore,
the tree amplitudes of the $B_s^0\to \rho^0 \phi$ channel from the factorizable emission diagrams Fig.~\ref{fig2}(a) and~\ref{fig2}(b) are also suppressed by the small Wilson coefficients $C_1+C_2/3$.
The dominant contributions are then from the penguin operators.
However as show in Eq.~(\ref{eq:b0r0phi}),
the $B_s^0\to \rho^0 \phi$ decay has no gluonic penguin amplitudes
because of the cancellations between the $u{\bar u}$ and $d{\bar d}$ component in the $\rho^0$ meson.
The only left parts are all electroweak penguin suppressed.
As a result,
the total branching ratio of the $B_s^0\to \rho^0 \phi$ is estimated to be small, at the order of $10^{-7}$.

For other two $B^{0}\to \rho^{0} \phi$ and $B^{+}\to \rho^{+} \phi$ decay channels controlled by $b\to d$ transitions,
the predicted branching ratios are much smaller than that of  $B_s^0\to \rho^0 \phi$ decay
due to the CKM-suppressed factor $|V_{td}/V_{ts}|^2\sim 0.05$.
The calculated ${\cal B}(B^{+}\to \rho^{+} \phi)=(0.013\pm 0.005)\times 10^{-6}$
and ${\cal B}(B^{0}\to \rho^{0} \phi)=(0.006 \pm 0.002)\times 10^{-6}$ in this work
are about half of the previous two-body results ${\cal B}(B^{+}\to \rho^{+} \phi)=(0.028^{+0.015}_{-0.013})\times 10^{-6}$~\cite{prd91-054033}
and ${\cal B}(B^{0}\to \rho^{0} \phi)=(0.013^{+0.007}_{-0.006})\times 10^{-6}$~\cite{prd91-054033}.
The main reason is that the additional higher power corrections related to the momenta fraction $x_B$ have been taken into account in the current work,
which has been ignored in Ref.~\cite{prd91-054033}.
Taking the $B^{0}\to \rho^{0} \phi$ decay as an example,
we have reexamined the branching fraction without the contributions from $x_B$:
${\cal B}(B^{0}\to \rho^{0} \phi)=0.01\times 10^{-6}$, which becomes similar to the previous two-body analysis.
The current experiments give the upper limits:
${\cal B}(B^+\to \rho^+\phi)<3.0 \times 10^{-6}$~\cite{pdg2020} and ${\cal B}(B^0\to \rho^0\phi)<3.3 \times 10^{-7}$~\cite{pdg2020} at $90\%$ C.L,
so more precise measurements are expected to differentiate these theoretical predictions.
Besides, under the isospin limit
the following relation among the $B^{0}\to \rho^{0} \phi$ and $B^{+}\to \rho^{+} \phi$ decays  is naively expected
\begin{eqnarray}
R=\frac{{\cal B}(B^{0}\to \rho^{0} \phi)}{{\cal B}(B^{+}\to \rho^{+} \phi)}\approx\frac{1}{2}\cdot \frac{\tau_{B^0}}{\tau_{B^+}}.
\end{eqnarray}
Our calculations basically agree with the relation given above and can be tested by the future experiments.

In the naive factorization approach,
the longitudinal polarizations are expected to dominate the branching ratios of charmless $B\to VV$ decays
according to the naive counting rules~\cite{plb622-63}
\begin{eqnarray}\label{eq:cr}
f_0\sim 1-\mathcal{O}(m_V^2/m_B^2),\quad f_{\parallel}\sim f_{\perp}\sim \mathcal{O}(m_V^2/m_B^2),
\end{eqnarray}
with $m_V$ being the vector meson mass.
In sharp contrast to these expectations, large transverse polarization of order $50\%$ is observed in the penguin dominated decays
$B\to K^*\rho$, $B\to K^*\phi$,
and $B^0_s\to \phi\phi$~\cite{prd85-072005,jhep05-026,prl91-201801,prd78-092008,prl107-261802,plb713-369},
which reflects that the counting rules given in Eq.~(\ref{eq:cr}) is violated
and poses an interesting challenge for the theory.
In order to interpret this large transverse polarization, a number of strategies have been proposed within or beyond the
SM~\cite{npb774-64,prd71-054025,prd70-054015,Cheng:2008gxa,Grossman:2003qi,Das:2004hq,Chen:2005mka,Yang:2004pm,
Kagan:2004uw,Beneke:2005we,Datta:2007qb,Colangelo:2004rd,Ladisa:2004bp,
Cheng:2004ru,
Bobeth:2014rra,Cheng:2010yd,Chen:2007qj,Chen:2005cx,Faessler:2007br,Chen:2006vs,Huang:2005qb,Baek:2005jk,Yang:2005tv,Alvarez:2004ci}.
In the PQCD approach,
the unexpected large transverse components are led by the penguin annihilation diagrams,
especially the $(S-P)(S+P)$ penguin annihilation, introduced by the QCD penguin operator $O_6$~\cite{prd71-054025},
which is originally introduced in Ref.~\cite{plb601-151}.

For the $B^{0 } \to \rho^{0 }\phi$, $B^{+} \to \rho^{+}\phi$ and $B^0_s \to \rho^0\phi$ decays,
the longitudinal polarization fractions are predicted to be as large as $90\%$,
which agree well with the previous PQCD calculations~\cite{prd76-074018,prd91-054033} and those from QCDF~\cite{prd80-114026}, SCET~\cite{prd96-073004}
and FAT~\cite{epjc77-333} within uncertainties.
The $B_{(s)} \to \rho\phi$  are  pure emission-type decays,
and the contributions  from the  chirally enhanced $(S-P)(S+P)$ penguin annihilation operator vanishes.
Besides
taking the $B^0_s \to \rho^0\phi$ decay as an example,
the dominant contributions are from the $3/2[C_7+C_8/3+C_9+C_{10}/3]F^{LL,h}_{e\phi}$ ($h=0, \parallel, \perp$)
induced by the electroweak penguin operators in the factorizable emission diagrams.
Compared with the longitudinal component $F^{LL,0}_{e\phi}$,
the transverse amplitudes $F^{LL,\|}_{e\phi}$ and $F^{LL, \bot}_{e\phi}$
are always highly suppressed by the factor $\omega_{\pi\pi}/m_{B_s}\approx m_\rho/m_{B_s}\approx 0.02$,
which leads to $f_0\sim 90\%$.

\subsection{$CP$-violating observables}

The direct $CP$ asymmetries with each helicity state ($\mathcal{A}_{0,\parallel,\perp}^{\text{CP}}$) of the four-body $B^0_{s} \to \rho^0\phi\to (\pi^+\pi^-)(K^+K^-)$ decay together with those summed over all helicity states ($\mathcal{A}_{\text{dir}}^{\text{CP}}$) are listed in  Table~\ref{tab:cpv}.
For comparison, we also present the updated results of the QCDF~\cite{prd80-114026}, SCET~\cite{prd96-073004}, FAT~\cite{epjc77-333}, and the PQCD~\cite{prd91-054033} predictions in two-body framework.
Meanwhile, the direct $CP$ asymmetry ${\cal A}^{\rm CP}$ of the $S$-wave decays $B_{(s)} \to [ VS, SV,SS]\to (\pi\pi)(K\bar K)$
are also displayed in Table \ref{tab:scp}.
The kinematics of the two-body decays is fixed,
while the amplitudes of the quasi-two-body decays depend on the invariant mass of the final-state pairs,
resulting in the differential distribution of direct $CP$ asymmetries.
The $CP$ asymmetry in the four-body framework is moderated by the finite width of the intermediate resonance appearing in the time-like form factor $F(\omega^2)$.
Hence, it is reasonable to see the differences of direct $CP$ asymmetries between the two-body and four-body frameworks in the PQCD approach as shown in Table ~\ref{tab:cpv}.

The direct $CP$ asymmetries of the two ${\bar b}\to {\bar d}{\bar s}s$ transition decays
$B^0\to \rho^0\phi\to(\pi^+\pi^-)(K^+K^-)$ and $B^+\to \rho^+\phi\to(\pi^+\pi^0)(K^+K^-)$
are naturally expected to be zero since only penguin operators work on these decays.
However,
the $B_s^0\to \rho^0\phi\to(\pi^+\pi^-)(K^+K^-)$ mode receives the additional tree contributions,
and  the interference between the tree and penguin amplitudes leads to the direct $CP$ asymmetry: ${\cal A}_{\rm dir}^{\rm CP}=(21.8^{+2.7}_{-3.3})\%$.
For the $S$-wave decays shown in Table \ref{tab:scp},
it is interesting to see that the predicted ${\cal A}^{\rm CP}$ of
the two double $S$-wave decays $B^0\to f_0f_0\to(\pi^+\pi^-)(K^+K^-)$ and $B_s^0\to f_0f_0\to(\pi^+\pi^-)(K^+K^-)$
are indeed quiet different.
As can be seen from the related numerical results in Table~\ref{tab:0sf0f0},
the fact is that for the $B_s^0\to f_0f_0\to(\pi^+\pi^-)(K^+K^-)$ decay,
the tree operators are highly suppressed by the CKM matrix elements $|V_{us}V^*_{ub}|$,
in comparison with $|V_{ts}V^*_{tb}|$ related to the penguin operators.
For $B^0\to f_0f_0\to(\pi^+\pi^-)(K^+K^-)$ decay,
both of $|V_{ud}V^*_{ub}|$ and $|V_{td}V^*_{tb}|$ are in the same order ($10^{-3}$),
which can strengthen the interference between the tree and penguin amplitudes.
Therefore,
the predicted ${\cal A}^{\rm CP}$ of the $B_s^0\to f_0f_0\to(\pi^+\pi^-)(K^+K^-)$
is much smaller than that of the $B^0\to f_0f_0\to(\pi^+\pi^-)(K^+K^-)$ decay.

As it is known that the direct $CP$ asymmetry depends on both the strong phase and the weak CKM phase.
In the SCET,
the large strong phase is only from the long-distance charming penguin at leading power and leading order.
In the QCDF and PQCD approaches, the strong phase comes from the hard spectator scattering and annihilation diagrams respectively.
So, the origins of strong phase are actually different in these three approaches,
which leads to different predictions of $\mathcal{A}_{\text{dir}}^{\text{CP}}(B_s^0\to \rho^0 \phi)$.
The forthcoming LHCb and Belle-II measurements for the direct $CP$ asymmetries can help us to examine these factorization approaches.

\begin{table}[t]
\caption{Direct $CP$ asymmetries (in units of $\%$) for the $B_s^0 \to \rho^0\phi\to(\pi^+\pi^-)(K^+K^-)$ decay compared with the previous predictions in the PQCD approach~\cite{prd91-054033}, the updated predictions in the QCDF~\cite{prd80-114026},  SCET~\cite{prd96-073004} and FAT~\cite{epjc77-333}.
The sources of the theoretical errors are the same as in Table~\ref{tab:brfour}.}
\label{tab:cpv}
\begin{ruledtabular}

\begin{tabular}[t]{lcccc}
Modes  & $\mathcal{A}_0^{\text{CP}}$ & $\mathcal{A}_{\parallel}^{\text{CP}}$& $\mathcal{A}_{\perp}^{\text{CP}}$ & $\mathcal{A}_{\text{dir}}^{\text{CP}}$\\
\hline
$B_s^0 \to \rho^0\phi\to(\pi^+\pi^-)(K^+K^-)$  &$29.4^{+0.4+0.9+2.9}_{-0.1-1.0-3.8}$&$-27.7^{+0.8+0.5+5.2}_{-1.1-0.4-5.6}$
                                                     &$-26.6^{+0.7+0.6+4.7}_{-1.0-0.5-4.9}$&$21.8^{+0.6+1.1+2.4}_{-0.1-1.1-3.1}$\\
PQCD~\cite{prd91-054033}  &$3.27^{+1.07}_{-1.19}$&&$-32.8^{+7.4}_{-5.8}$&$-4.3^{+1.5}_{-1.2}$\\
QCDF \cite{prd80-114026}   & $\cdots$  &$\cdots$  &$\cdots$&$83^{+10}_{-36}$\\
SCET \cite{prd96-073004}  & $\cdots$  &$\cdots$  &$\cdots$&$0$\\
FAT   \cite{epjc77-333} & $\cdots$  &$\cdots$  &$\cdots$ &$0$\\
\end{tabular}
\end{ruledtabular}
\end{table}

\begin{table}[!htbh]
\caption{PQCD predictions for the  direct $CP$ asymmetries ${\cal A}^{\rm CP}$ (in units of $\%$) of  the $B_{(s)}\rightarrow [VS,SV,SS]\to (\pi\pi)(K \bar K)$ decays, with $S=f_0(980)$ and $V=\rho,\phi$.
The sources of the theoretical errors are the same as in Table~\ref{tab:brfour}.}
\label{tab:scp}
\begin{tabular*}{12cm}{@{\extracolsep{\fill}}ll} \hline\hline
decay modes                                           &{\rm PQCD predictions}                                       \\ \hline
$B^+\to \rho^+f_0\to (\pi^+\pi^0)(K^+K^-)$          &$-8.8_{-0.9-0.8-0.9}^{+0.8+1.2+0.9}$                     \\
$B^0\to \rho^0 f_0\to (\pi^+\pi^-)(K^+K^-)$          &$-17.6_{-6.2-14.9-23.4}^{+0.0+13.7+26.2}$                     \\
$B^0_s\to \rho^0 f_0\to (\pi^+\pi^-)(K^+K^-)$          &$21.3_{-0.6-3.9-3.8}^{+1.4+3.3+2.9}$                     \\
$B^0\to f_0 \phi\to   (\pi^+\pi^-)(K^+K^-)$          &$0.0$                     \\
$B^0_s \to f_0 \phi\to   (\pi^+\pi^-)(K^+K^-)$          &$5.4_{-0.0-2.0-4.8}^{+2.9+3.9+1.6}$                     \\
$B^0\to f_0 f_0 \to   (\pi^+\pi^-)(K^+K^-)$          &$-79.4_{-2.2-1.1-2.3}^{+7.7+25.3+4.4}$                     \\
$B^0_s\to f_0 f_0 \to   (\pi^+\pi^-)(K^+K^-)$          &$-0.06_{-3.41-5.50-3.22}^{+0.00+0.00+0.00}$                     \\
\hline\hline
\end{tabular*}
\end{table}

\begin{table}[!htbh]
\caption{Branching ratios of the double $S$-wave four-body decays $B^0\to f_0 f_0 \to   (\pi^+\pi^-)(K^+K^-)$
and $B^0_s\to f_0 f_0 \to   (\pi^+\pi^-)(K^+K^-)$ from different topology diagrams.}
\label{tab:0sf0f0}
\begin{tabular*}{12cm}{@{\extracolsep{\fill}}lll} \hline\hline
Decay modes                                           &{\rm Tree contributions}             &{\rm Penguin contributions }                        \\ \hline
$B^0\to f_0 f_0 \to   (\pi^+\pi^-)(K^+K^-)$          &$0.78\times 10^{-10}$   &$1.12\times 10^{-10}$                   \\
$B^0_s\to f_0 f_0 \to   (\pi^+\pi^-)(K^+K^-)$          &$2.22\times 10^{-11}$    &$7.12\times 10^{-8}$                  \\
\hline\hline
\end{tabular*}
\end{table}

\begin{table}[t]
\caption{PQCD predictions for the TPAs ($\%$) of the four-body $B_{(s)} \to \rho\phi \to (\pi\pi)(K {\bar K})$ decays.
The sources of theoretical errors are the same as in Table~\ref{tab:brfour} but added in quadrature.}
\label{tab:tpas}
\begin{center}
\begin{threeparttable}
\begin{tabular}{l|c|c|c|c|c|c}
\hline\hline
\multicolumn{1}{c|}{}  &\multicolumn{6}{c}{ $\text{TPAs}$-1}   \cr\cline{2-7}
{Modes} &$\mathcal{A}_{T}^1$&$\bar{\mathcal{A}}_{T}^1$&$\mathcal{A}_{\text{T-true}}^1$ &$\mathcal{A}_{\text{T-fake}}^1$&$\mathcal{A}_{\text{T-True}}^{(1)\text{ave}}$ &$\mathcal{A}_{\text{T-fake}}^{(1)\text{ave}}$  \cr \hline
$B^+ \to \rho^+\phi\to (\pi^+\pi^0)(K^+K^-)$  &$-20.92^{+6.26}_{-2.80}$&$20.92^{+2.80}_{-6.26}$&$0$
                                             &$-20.92^{+6.26}_{-2.80}$&$0$&$-20.92_{-2.80}^{+6.26} $\\ \hline

$B^0 \to \rho^0\phi\to (\pi^+\pi^-)(K^+K^-)$ &$-20.92^{+6.26}_{-2.80}$&$20.92^{+2.80}_{-6.26}$&$0$
                                             &$-20.92^{+6.26}_{-2.80}$&$0$&$-20.92_{-2.80}^{+6.26} $\\ \hline

$B_s^0 \to \rho^0\phi\to (\pi^+\pi^-)(K^+K^-)$  &$-23.53^{+1.05}_{-0.62}$&$3.06^{+3.22}_{-3.11}$&$-10.23^{+1.73}_{-1.56}$
                                             &$-13.29^{+1.79}_{-1.45}$&$-11.42^{+1.82}_{-1.58}$&$-14.20^{+1.51}_{-1.25}$\\ \hline
\hline
\multicolumn{1}{c|}{}  &\multicolumn{6}{|c}{$\text{TPAs}$-2}   \cr\cline{2-7}
{Modes} &$\mathcal{A}_{T}^2$&$\bar{\mathcal{A}}_{T}^2$ &$\mathcal{A}_{\text{T-true}}^2$ &$\mathcal{A}_{\text{T-fake}}^2$&$\mathcal{A}_{\text{T-True}}^{(2)\text{ave}}$ &$\mathcal{A}_{\text{T-fake}}^{(2)\text{ave}}$  \cr \hline
$B^+ \to \rho^+\phi\to (\pi^+\pi^0)(K^+K^-)$ &$1.02^{+1.12}_{-0.54}$&$-1.02^{-1.12}_{+0.54}$&$0$
                                             &$1.02^{+1.12}_{-0.54}$&$0$&$1.02^{+1.12}_{-0.54}$\\ \hline

$B^0 \to \rho^0\phi\to (\pi^+\pi^-)(K^+K^-)$ &$1.02^{+1.12}_{-0.54}$&$-1.02^{-1.12}_{+0.54}$&$0$
                                             &$1.02^{+1.12}_{-0.54}$&$0$&$1.02^{+1.12}_{-0.54}$\\ \hline

$B_s^0 \to \rho^0\phi\to (\pi^+\pi^-)(K^+K^-)$ &$-4.91^{+0.30}_{-0.24}$&$-0.08^{+0.73}_{-0.76}$&$-2.50^{+0.43}_{-0.37}$
                                             &$-2.42^{+0.25}_{-0.32}$&$-2.72^{+0.44}_{-0.35}$&$-2.64^{+0.36}_{-0.27}$\\ \hline

\hline\hline
\end{tabular}
\end{threeparttable}
\end{center}
\end{table}

The PQCD predictions for the ``true" and ``fake" TPAs of the $B_{(s)}\to\rho\phi\to (\pi\pi)(K \bar K)$ decays are collected in Table~\ref{tab:tpas}.
As mentioned previously,
the averaged asymmetries ${\cal A}^{1(2),\text{ave}}_{\text{T-true}}({\cal A}^{1(2),\text{ave}}_{\text{T-fake}})$
are usually not equal to the so-called ``true" (``fake") asymmetries ${\cal A}^{1(2)}_{\text{T-true}}({\cal A}^{1(2)}_{\text{T-fake}})$.
They become equal only in the absence of direct $CP$ violation in the total rates, namely $\mathcal{D}=\bar{\mathcal{D}}$,
such as  the $B^0\to \rho^0\phi \to (\pi^+\pi^-)(K^+K^-)$ and $B^+\to \rho^+\phi \to (\pi^+\pi^0)(K^+K^-)$ decays.

For the two pure penguin decays $B^0\to \rho^0\phi \to (\pi^+\pi^-)(K^+K^-)$ and $B^+\to \rho^+\phi \to (\pi^+\pi^0)(K^+K^-)$,
each helicity amplitude involves the same single weak phase in the SM,
resulting in $A^i_T=-{\bar A}^i_T$ due to the vanishing weak phase difference.
The ``true" TPAs for these two decay channels are thus expected to be zero.
If such asymmetries are observed experimentally,
it is probably a signal of new physics.
However,
since the $B^0_s \to \rho^0\phi\to (\pi^+\pi^-)(K^+K^-)$ decay can receive extra tree contributions,
the magnitude of the calculated ``true" TPA can exceed ten percent,
which is expected to be tested  by the future experiments.
As ``fake" TPAs are due to strong phases and require no weak phase difference,
the predicted ${\cal A}^{1(2)}_{\text{T-fake}}$ and ${\cal A}^{1(2), \text{ave}}_{\text{T-fake}}$are usually  nonzero for all considered decays.
The predicted large ``fake" asymmetry ${\cal A}^{1}_{\text{T-fake}}=(-20.92^{+6.26}_{-2.80})\%$ of the $B^0\to \rho^0\phi \to (\pi^+\pi^-)(K^+K^-)$ decay
simply reflect the final-state strong phases.

As usual,
the decay amplitude associated with transverse polarization $A_{\parallel}$ is smaller than that for longitudinal polarization $A_0$ in the SM within factorization.
This indicates that $\mathcal{A}_\text{T}^2$ is power suppressed relative to $\mathcal{A}_\text{T}^1$.
Meanwhile,
the smallness of $\mathcal{A}_\text{T}^2$  is also attributed to the suppression from the strong phase difference between the perpendicular and parallel polarization amplitudes, which is supported by the previous PQCD calculations~\cite{prd91-054033}.
An observation of $\mathcal{A}_\text{T}^2$ with large values can signify physics beyond the SM.
All these PQCD predictions can be tested in the near future.

\section{Conclusion}\label{sec:5}
By employing the perturbative QCD factorization approach,
we have systematically investigated the four-body decays $B_{(s)} \to (\pi\pi)(K {\bar K})$ under the quasi-two-body approximation,
in which the $\pi\pi$ and $K \bar K$ invariant-mass spectrum are dominated by the vector resonances $\rho^0$ and $\phi$, respectively.
The scalar resonance $f_0(980)$ is also contributed in the selected $\pi\pi$ and $K {\bar K}$ invariant-mass ranges.
The strong dynamics associated with the hadronization of the final state meson pairs is parametrized into the non-perturbative two-meson DAs,
which include both resonant and nonresonant contributions and have been established in three-body $B$ meson decays.
With the two-meson DAs,
the branching ratios, polarization fractions, direct $CP$ asymmetries, and the triple product asymmetries
of the four-body decays $B_{(s)} \to [\rho\phi, \rho f_0, f_0\phi, f_0f_0]\to (\pi\pi)(K \bar K)$ have been examined.

Under the narrow width approximation equation,
the two-body $B_{(s)} \rightarrow \rho\phi$ branching ratios have been extracted from
the results for the four-body decays $B_{(s)} \to\rho\phi\to (\pi\pi)(K \bar K)$.
We also presented the polarization fractions of the related four-body decays.
The obtained two-body branching ratio ${\cal B}(B^0_{s} \to\rho^0\phi)$ is consistent well with the previous two-body PQCD prediction
and the current experimental data within errors.
The calculated large  longitudinal polarization fractions $f_0 \sim 90\%$ of the $B_{(s)} \rightarrow \rho\phi$ decay modes also agree well
with the theoretical predictions from the previous PQCD, QCDF, SCET and FAT approaches.

We  calculated the direct $CP$ asymmetries and TPAs of the four-body $B_{(s)} \to (\pi\pi)(K\bar K)$ decays.
For the two pure penguin $B^0\to \rho^0\phi\to(\pi^+\pi^-)(K^+K^-)$ and $B^+\to \rho^+\phi\to(\pi^+\pi^0)(K^+K^-)$ decays,
both the direct $CP$ asymmetries and ``true" TPAs are naturally expected to be zero in the SM due to the vanishing weak phase difference.
While for $B_s^0\to \rho^0\phi\to(\pi^+\pi^-)(K^+K^-)$ channel,
the magnitude of the ${\cal A}_{\rm dir}^{\rm CP}$ and ${\cal A}^1_{\text{T-true}}$
can exceed $20\%$ and $10\%$  respectively,
which are expected to be confronted with the future experiments.
The ``fake" TPAs requiring no weak phase difference are usually nonzero for all decay channels.
The predicted sizable  ${\cal A}^{1}_{\text{T-fake}}=(-20.92^{+6.26}_{-2.80})\%$
of the $B^0\to \rho^0\phi\to(\pi^+\pi^-)(K^+K^-)$  decay
simply reflects the importance of the strong final-state interactions.

\begin{acknowledgments}
Many thanks to H.n.~Li for valuable discussions.
This work was supported by the National Natural Science Foundation of China under the No.~12075086, No.~12105028.
ZR is supported in part by the Natural Science Foundation of Hebei Province under Grant No.~A2021209002  and No.~A2019209449.

\end{acknowledgments}

\appendix
\section{$S$-wave Decay amplitudes}\label{samp}
According to Eq.~(\ref{f0mix}),
the total decay amplitudes of the $S$-wave channels can be divided  into the
$n {\bar n}=\frac{1}{\sqrt{2}}(u{\bar u}+d{\bar d})$ and $s{\bar s}$ components,
\begin{itemize}
\item[]

$\bullet$ $ B_{(s)} \to [\rho f_0, f_0 \phi] \to(\pi\pi)(K{\bar K})$ decay modes

\begin{eqnarray}
A(B^+ \to \rho^+ f_n\to(\pi^+\pi^0)(K^+K^-))&=& \frac{G_F} {2}V_{ub}^*V_{ud}\Big[\left (C_2+\frac{C_1}{3}\right)(F^{LL}_{ef_n}+F^{LL}_{af_n}+F^{LL}_{a\rho})\non
&+&C_1\left ( M^{LL}_{ef_n}+M^{LL}_{af_n}+M^{LL}_{a\rho}\right )+C_2M^{LL}_{e\rho}\Big]\non
&-& \frac{G_F} {2}V_{tb}^*V_{td}\Big[\left (C_4+\frac{C_3}{3}+C_{10}+\frac{C_9}{3}\right)(F^{LL}_{ef_n}+F^{LL}_{af_n}+F^{LL}_{a\rho})\non
&+& \left(C_6+\frac{C_5}{3}+C_8+\frac{C_7}{3} \right)(F^{SP}_{af_n}+F^{SP}_{a\rho})+\left( 2C_6+\frac{C_{8}}{2}\right)M^{SP}_{e\rho} \non
&+&\left(C_6+\frac{C_5}{3}-\frac{C_8}{2}-\frac{C_{7}}{6} \right)F^{SP}_{e\rho}+\left( C_5-\frac{C_7}{2}\right )M^{LR}_{e\rho}\non
&+&\left( C_3+C_9\right )(M^{LL}_{ef_n}+M^{LL}_{af_n}+M^{LL}_{a\rho})\non
&+& \left( C_5+C_7\right )(M^{LR}_{ef_n}+M^{LR}_{af_n}+M^{LR}_{a\rho})\non
&+&\left(C_3+2C_4-\frac{C_9}{2}+\frac{C_{10}}{2} \right )M^{LL}_{e\rho}\Big],\label{f0rp}
\\\non
A(B^+ \to \rho^+f_s\to(\pi^+\pi^0)(K^+K^-))&=&
- \frac{G_F} {\sqrt{2}}V_{tb}^*V_{td}\Big[\left (C_4-\frac{C_{10}}{2}\right)M^{LL}_{e\rho}+\left (C_6-\frac{C_{8}}{2}\right)M^{SP}_{e\rho}\Big],
\\\non
2A(B^0 \to \rho^0f_n\to(\pi^+\pi^-)(K^+K^-))&=& \frac{G_F} {\sqrt{2}}V_{ub}^*V_{ud}\Big[\left (C_1+\frac{C_2}{3}\right)(F^{LL}_{ef_n}+F^{LL}_{af_n}+F^{LL}_{a\rho})\non
&+&C_2\left ( M^{LL}_{ef_n}+M^{LL}_{af_n}+M^{LL}_{e\rho}+M^{LL}_{a\rho}\right )\Big]\non
&-& \frac{G_F} {\sqrt{2}}V_{tb}^*V_{td}\Big[\left (-C_4-\frac{C_3}{3}+\frac{3C_7}{2}+\frac{C_{8}}{2}+\frac{3C_9}{2}
+\frac{C_{10}}{2}+\frac{C_{10}}{2}+\frac{C_9}{6}\right)\non
&\times& \left(F^{LL}_{ef_n}+F^{LL}_{af_n}+F^{LL}_{a\rho}\right) +\left(-C_6-\frac{C_5}{3}+\frac{C_8}{2}+\frac{C_{7}}{6} \right)(F^{SP}_{e\rho}+F^{SP}_{af_n}+F^{SP}_{a\rho})\non
&-&\left( C_3-\frac{C_9}{2}-\frac{3C_{10}}{2}\right )(M^{LL}_{ef_n}+M^{LL}_{af_n}+M^{LL}_{a\rho})-\left( 2C_6+\frac{C_{8}}{2}\right)M^{SP}_{e\rho} \non
&-& \left(C_3+2C_4-\frac{C_9}{2}+\frac{C_{10}}{2} \right )M^{LL}_{e\rho}+\frac{3C_{8}}{2} (M^{SP}_{ef_n}+M^{SP}_{af_n}+M^{SP}_{a\rho})\non
&-&\left( C_5-\frac{C_7}{2}\right )(M^{LR}_{ef_n}+M^{LR}_{af_n}+M^{LR}_{e\rho}+M^{LR}_{a\rho})\Big],\label{f0r0}
\\\non
A(B^0 \to \rho^0f_s\to(\pi^+\pi^-)(K^+K^-))&=&
- \frac{G_F} {2}V_{tb}^*V_{td}\Big[\left (-C_4+\frac{C_{10}}{2}\right)M^{LL}_{e\rho}+\left (-C_6+\frac{C_{8}}{2}\right)M^{SP}_{e\rho}\Big],
\end{eqnarray}
\begin{eqnarray}
2A(B_s^0 \to \rho^0f_n\to(\pi^+\pi^-)(K^+K^-))&=&\frac{G_F} {\sqrt{2}}V_{ub}^*V_{us}\Big[\left (C_1+\frac{C_2}{3}\right)(F^{LL}_{af_n}+F^{LL}_{a\rho})
+C_2(M^{LL}_{af_n}+M^{LL}_{a\rho})\Big]\non
&-& \frac{G_F} {\sqrt{2}}V_{tb}^*V_{ts}\Big[\left (\frac{3C_{7}}{2}+\frac{C_8}{2}+\frac{3C_{9}}{2}+\frac{C_{10}}{2}\right)(F^{LL}_{af_n}+F^{LL}_{a\rho})\non
&+&\frac{3C_{10}}{2}(M^{LL}_{af_n}+M^{LL}_{a\rho})+\frac{3C_{8}}{2}(M^{SP}_{af_n}+M^{SP}_{a\rho})\Big],
\\\non
A(B_s^0 \to \rho^0f_s\to(\pi^+\pi^-)(K^+K^-))&=&\frac{G_F} {2}\Big\{V_{ub}^*V_{us}\Big[\left(C_1+\frac{C_2}{3}\right )F^{LL}_{ef_s}
+C_2M^{LL}_{ef_s}\Big ]  \non
 &-&V_{tb}^*V_{ts}\Big [  \frac{3}{2}\left(C_7+\frac{C_8}{3}+C_9+\frac{C_{10}}{3} \right )F^{LL}_{ef_s}  \non
 &+&  \frac{3C_{10}}{2}M^{LL}_{ef_s}+\frac{3C_{8}}{2}M^{SP}_{ef_s}\Big ]\Big\},\label{eq:b0r0f0}
\\\non
A(B^0 \to f_n \phi\to(\pi^+\pi^-)(K^+K^-))&=& - \frac{G_F} {2}V_{tb}^*V_{td}\Big[
\left(C_3+\frac{C_4}{3}+C_5+\frac{C_6}{3}-\frac{C_7}{2}-\frac{C_{8}}{6}- \frac{C_9}{2}-\frac{C_{10}}{6} \right )F^{LL,h}_{ef_n}\non
&+&\left(2C_4+\frac{C_{10}}{2} \right )M^{LL}_{ef_n}+\left( 2C_6+\frac{C_{8}}{2}\right)M^{SP}_{ef_n}\Big],
\\\non
A(B^0 \to f_s \phi\to(\pi^+\pi^-)(K^+K^-))&=& - \frac{G_F} {\sqrt{2}}V_{tb}^*V_{td}\Big[
\left(C_3+\frac{C_4}{3}+C_5+\frac{C_6}{3}-\frac{C_7}{2}-\frac{C_{8}}{6}- \frac{C_9}{2}-\frac{C_{10}}{6} \right )\non
&\times& (F^{LL,h}_{af_s}+F^{LL,h}_{a\phi})+\left(2C_4+\frac{C_{10}}{2} \right )(M^{LL}_{af_s}+M^{LL}_{a\phi})\non
&+&\left( 2C_6+\frac{C_{8}}{2}\right)(M^{SP}_{af_s}+M^{SP}_{a\phi})\Big],
\\\non
A(B_s^0 \to f_n \phi\to(\pi^+\pi^-)(K^+K^-))&=& \frac{G_F} {2}V_{ub}^*V_{us}\Big[C_2M^{LL}_{e\phi}\Big]\non
&-& \frac{G_F} {2}V_{tb}^*V_{ts}\Big[ \left(2C_4+\frac{C_{10}}{2} \right )M^{LL}_{e\phi}+\left( 2C_6+\frac{C_{8}}{2}\right)M^{SP}_{e\phi}\Big],
\\\non
A(B_s^0 \to f_s\phi\to(\pi^+\pi^-)(K^+K^-))&=&-\frac{G_F} {\sqrt{2}}V_{tb}^*V_{ts}\Big [\left (C_5-\frac{C_7}{2}\right )\left (M^{LR}_{e\phi}+M^{LR}_{a\phi}+M^{LR}_{ef_s}+M^{LR}_{af_s}\right)\non
&+&\frac{4}{3} \left (C_3+C_4-\frac{C_9}{2}-\frac{C_{10}}{2} \right ) \left (F^{LL}_{ef_s}+F^{LL}_{a\phi}+F^{LL}_{af_s}\right )\non
&+& \left(C_6-\frac{C_8}{2}\right)\left(M^{SP}_{ef_s}+M^{SP}_{af_s}+M^{SP}_{e\phi}+M^{SP}_{a\phi}\right)\non
&+& \left ( C_3+C_4-\frac{C_9}{2}-\frac{C_{10}}{2} \right ) \left (M^{LL}_{e\phi}+M^{LL}_{a\phi}+M^{LL}_{ef_s}+M^{LL}_{af_s} \right )\non
&+&\left (C_5+\frac{C_6}{3}-\frac{C_7}{2}-\frac{C_{8}}{6}\right ) \left (F^{LR}_{ef_s}+F^{LR}_{a\phi}+F^{LR}_{af_s}\right )\non
&+&\left ( C_6+\frac{C_5}{3}-\frac{C_8}{2}-\frac{C_{7}}{6}\right)\left (F^{SP}_{e\phi}+F^{SP}_{a\phi}+F^{SP}_{af_s}\right )\Big ].\label{eq:b0f0phi}
\end{eqnarray}

The decay amplitudes for the physical states are then
\begin{eqnarray}
A(B_{(s)} \to [\rho f_0, f_0\phi]\to(\pi\pi)(K{\bar K}))&=&A(B_{(s)} \to [\rho f_n, f_n\phi]\to(\pi\pi)(K{\bar K}))\sin \theta\non
                      &+&A(B_{(s)} \to [\rho f_s, f_s\phi]\to(\pi\pi)(K{\bar K}))\cos \theta.
\end{eqnarray}

\item[]
$\bullet$ $B \to f_0 f_0\to(\pi\pi)(K{\bar K})$ decay modes
\begin{eqnarray}
2A(B^0 \to f_n f_n\to(\pi^+\pi^-)(K^+K^-))&=& \frac{G_F} {\sqrt{2}}V_{ub}^*V_{ud}\Big[\left (C_1+\frac{C_2}{3}\right)(F^{LL}_{af_n})
+C_2\left ( M^{LL}_{ef_n}+M^{LL}_{af_n}\right )\Big]\non
&-& \frac{G_F} {\sqrt{2}}V_{tb}^*V_{td}\Big[ \Big(2C_3+\frac{2C_4}{3}+C_4+\frac{C_3}{3}+2C_5+\frac{2C_6}{3}+\frac{C_7}{2}+\frac{C_8}{6}\non
&+&\frac{C_{9}}{2}+\frac{C_{10}}{6}-\frac{C_{10}}{2}-\frac{C_{9}}{6} \Big)F^{LL}_{af_n}+\left(C_6+\frac{C_5}{3}-\frac{C_8}{2}-\frac{C_{7}}{6} \right)\left(F^{SP}_{ef_n}+F^{SP}_{af_n}\right)\non
&+&\left(C_3+2C_4-\frac{C_9}{2}+\frac{C_{10}}{2} \right )\left(M^{LL}_{ef_n}+M^{LL}_{af_n}\right)+\left( C_5-\frac{C_7}{2}\right )\left(M^{LR}_{ef_n}+M^{LR}_{af_n}\right)\non
&+& \left( 2C_6+\frac{C_{8}}{2}\right)\left(M^{SP}_{ef_n}+M^{SP}_{af_n}\right) \Big]\non
&+&[(f_n\to K^+K^-) \leftrightarrow (f_n\to \pi^+\pi^-)],
\\\non
A(B^0 \to f_n f_s\to(\pi^+\pi^-)(K^+K^-))&=& -\frac{G_F} {2}V_{tb}^*V_{td}\Big[ \left(C_4-\frac{C_{10}}{2} \right )M^{LL}_{ef_n}+\left( C_6-\frac{C_{8}}{2}\right)M^{SP}_{ef_n}\Big],
\\\non
A(B^0 \to f_s f_s\to(\pi^+\pi^-)(K^+K^-))&=& -\frac{G_F} {\sqrt{2}}V_{tb}^*V_{td}\Big[ \Big(C_3+\frac{C_4}{3}+C_5+\frac{C_6}{3} -\frac{C_7}{2}-\frac{C_8}{6}
-\frac{C_{9}}{2}-\frac{C_{10}}{6}\Big)F^{LL}_{af_s}\non
&+&\left(C_4-\frac{C_{10}}{2} \right )M^{LL}_{af_s}+\left( C_6-\frac{C_{8}}{2}\right)M^{SP}_{af_s}\Big]\non
&+&[(f_s\to K^+K^-) \leftrightarrow (f_s\to \pi^+\pi^-)],
\non\\
2A(B_s^0 \to f_n f_n\to(\pi^+\pi^-)(K^+K^-))&=& \frac{G_F} {\sqrt{2}}V_{ub}^*V_{us}\Big[\left (C_1+\frac{C_2}{3}\right)F^{LL}_{af_n}
+C_2M^{LL}_{af_n}\Big]\non
&-& \frac{G_F} {\sqrt{2}}V_{tb}^*V_{ts}\Big[ \left(2C_3+\frac{2C_4}{3}+2C_5+\frac{2C_6}{3}+\frac{C_7}{2}+\frac{C_8}{6}+\frac{C_{9}}{2}+\frac{C_{10}}{6}\right)F^{LL}_{af_n}\non
&+&\left(2C_4+\frac{C_{10}}{2} \right )M^{LL}_{af_n}+\left( 2C_6+\frac{C_{8}}{2}\right)M^{SP}_{af_n}\Big]\non
&+&[(f_n\to K^+K^-) \leftrightarrow (f_n\to \pi^+\pi^-)],
\\\non
A(B_s^0 \to f_n f_s\to(\pi^+\pi^-)(K^+K^-))&=& \frac{G_F} {2}V_{ub}^*V_{us}\Big[C_2M^{LL}_{ef_s}\Big]\non
&-& \frac{G_F} {2}V_{tb}^*V_{ts}\Big[ \left(2C_4+\frac{C_{10}}{2} \right )M^{LL}_{ef_s}+\left( 2C_6+\frac{C_{8}}{2}\right)M^{SP}_{ef_s}\Big],
\\\non
A(B_s^0 \to f_s f_s\to(\pi^+\pi^-)(K^+K^-))&=& -\frac{G_F} {\sqrt{2}}V_{tb}^*V_{ts}\Big[ \Big(C_3+\frac{C_4}{3}+C_4+\frac{C_3}{3}-\frac{C_{9}}{2}-\frac{C_{10}}{6}-\frac{C_{10}}{2}-\frac{C_{9}}{6} \Big)F^{LL}_{af_s}\non
&+&\left(C_6+\frac{C_5}{3}-\frac{C_8}{2}-\frac{C_{7}}{6} \right)\left(F^{SP}_{ef_s}+F^{SP}_{af_s}\right)\non
&+&\left(C_3+C_4-\frac{C_9}{2}-\frac{C_{10}}{2} \right )\left(M^{LL}_{ef_s}+M^{LL}_{af_s}\right)+\left( C_5-\frac{C_7}{2}\right )\left(M^{LR}_{ef_s}+M^{LR}_{af_s}\right)\non
&+& \left( C_6-\frac{C_{8}}{2}\right)\left(M^{SP}_{ef_s}+M^{SP}_{af_s}\right) \Big]\non
&+&[(f_s\to K^+K^-) \leftrightarrow (f_s\to \pi^+\pi^-)].
\end{eqnarray}

The decay amplitudes for the physical states are then
\begin{eqnarray}
A(B_{(s)} \to f_0 f_0\to(\pi\pi)(K{\bar K}))&=&A(B_{(s)} \to f_n f_n\to(\pi\pi)(K{\bar K}))(\sin \theta)^2\non
&+&A(B_{(s)} \to f_n f_s\to(\pi\pi)(K{\bar K}))\sin 2\theta\non
&+&A(B_{(s)} \to f_s f_s\to(\pi\pi)(K{\bar K}))(\cos \theta)^2.
\end{eqnarray}

\end{itemize}

\section{Two-meson distribution amlitudes}\label{TMDAs}
The $S$-wave two-meson DAs can be written as~\cite{prd91-094024} ($h_1h_2=\pi\pi,K{\bar K}$),
\begin{eqnarray}\label{swave}
\Phi_{(h_1h_2)_S}(z,\omega)=\frac{1}{\sqrt{2N_c}}[{p\hspace{-1.5truemm}/}\phi^0_{(h_1h_2)_S}(z,\omega^2)+
\omega\phi^s_{(h_1h_2)_S}(z,\omega^2)+\omega({n\hspace{-2.0truemm}/}{v\hspace{-2.0truemm}/}-1)\phi^t_{(h_1h_2)_S}(z,\omega^2)],
\end{eqnarray}
in which $N_c$ is the number of colors, and the asymptotic forms of the various twists DAs are parametrized as~\cite{MP,MT01,MT02,MT03}
\begin{eqnarray}
\phi^0_{(h_1h_2)_S}(z,\omega^2)&=&\frac{9F_S(\omega^2)}{\sqrt{2N_c}}a_{h_1h_2}z(1-z)(1-2z),\label{eq:phis0}\\
\phi^s_{(h_1h_2)_S}(z,\omega^2)&=&\frac{F_S(\omega^2)}{2\sqrt{2N_c}},\\
\phi^t_{(h_1h_2)_S}(z,\omega^2)&=&\frac{F_S(\omega^2)}{2\sqrt{2N_c}}(1-2z),
\end{eqnarray}
with the time-like scalar form factor $F_S(\omega^2)$.
The Gegenbauer moments $a_{h_1h_2}$  in Eq.~(\ref{eq:phis0}) are adopted the same values as those determined in Refs.~\cite{epjc76-675,epjc79-792}:
$a_{\pi\pi}=0.20\pm 0.20$~\cite{epjc76-675}, $a_{KK}=0.80\pm 0.16$~\cite{epjc79-792}.

The elastic rescattering effects in the final-state meson pair can usually be absorbed into the time-like form factor
$F(\omega^2)$ according to the Watson theorem~\cite{pr88-1163}.
For the scalar resonance $f_0(980)$, its pole mass is very close to the $K{\bar K}$ threshold,
which can have strong influence on the resonance shape.
In the present work,
we follow Refs.~\cite{prd89-092006,prd90-012003} to employ the widely used Flatt\'e model suggested by D.V.~Bugg~\cite{prd78-074023},
\begin{eqnarray}
F_S(\omega^2)=\frac{m_{f_0(980)}^2}{m_{f_0(980)}^2-\omega^2-im_{f_0(980)}(g_{\pi\pi}\rho_{\pi\pi}+g_{KK}\rho_{KK}F^2_{KK})}\;,
\end{eqnarray}
with the two phase space factors $\rho_{\pi\pi}$ and $\rho_{KK}$~\cite{prd87-052001,prd89-092006,plb63-228}
\begin{eqnarray}
\rho_{\pi\pi}=\frac23\sqrt{1-\frac{4m^2_{\pi^\pm}}{\omega^2}}
 +\frac13\sqrt{1-\frac{4m^2_{\pi^0}}{\omega^2}},\quad
\rho_{KK}=\frac12\sqrt{1-\frac{4m^2_{K^\pm}}{\omega^2}}
 +\frac12\sqrt{1-\frac{4m^2_{K^0}}{\omega^2}}.
\end{eqnarray}
The $g_{\pi\pi}=0.167$ GeV and $g_{KK}=3.47g_{\pi\pi}$~\cite{prd89-092006,prd90-012003} are coupling constants,
describing the $f_0$ decay into the final states $\pi^+\pi^-$ and $K^+K^-$, respectively.
The exponential factor $F_{KK}=e^{-\alpha q_K^2}$
is introduced above the $K\bar{K}$ threshold to reduce the $\rho_{KK}$ factor as invariant mass increases,
where $q_k$ is the momentum of the kaon in the $K\bar{K}$ rest frame and $\alpha=2.0\pm 0.25$ GeV$^{-2}$~\cite{prd89-092006,prd78-074023}.

The corresponding $P$-wave two-meson DAs related to both longitudinal and transverse polarizations are decomposed,
up to the twist 3, into~\cite{prd98-113003}:
\begin{eqnarray}
\Phi_{(h_1h_2)_P}^{L}(z,\zeta,\omega)&=&\frac{1}{\sqrt{2N_c}} \left [{ \omega \epsilon\hspace{-1.5truemm}/_p  }\phi_{(h_1h_2)_P}^0(z,\omega^2)+\omega\phi_{(h_1h_2)_P}^s(z,\omega^2)
+\frac{{p\hspace{-1.5truemm}/}_1{p\hspace{-1.5truemm}/}_2
  -{p\hspace{-1.5truemm}/}_2{p\hspace{-1.5truemm}/}_1}{\omega(2\zeta-1)}\phi_{(h_1h_2)_P}^t(z,\omega^2) \right ] (2\zeta-1)\;,\label{pwavel}\\
\Phi_{(h_1h_2)_P}^{T}(z,\zeta,\omega)&=&\frac{1}{\sqrt{2N_c}}
\Big [\gamma_5{\epsilon\hspace{-1.5truemm}/}_{T}{ p \hspace{-1.5truemm}/ } \phi_{(h_1h_2)_P}^T(z,\omega^2)
+\omega \gamma_5{\epsilon\hspace{-1.5truemm}/}_{T} \phi_{(h_1h_2)_P}^a(z,\omega^2)+ i\omega\frac{\epsilon^{\mu\nu\rho\sigma}\gamma_{\mu}
\epsilon_{T\nu}p_{\rho}n_{-\sigma}}{p\cdot n_-} \phi_{(h_1h_2)_P}^v(z,\omega^2) \Big ]\non
&&\cdot \sqrt{\zeta(1-\zeta)+\alpha_1}\label{pwavet}\;.
\end{eqnarray}
The various twist DAs $\phi_{(h_1h_2)_P}^i$  in the above equations can be expanded in terms of the Gegenbauer polynomials:
\begin{eqnarray}
\phi_{\pi\pi}^0(z,\omega^2)&=&\frac{3F_{\pi\pi}^{\parallel}(\omega^2)}{\sqrt{2N_c}}z(1-z)\left[1
+a^0_{2\rho}\frac{3}{2}(5(1-2z)^2-1)\right] \;,\\
\phi_{\pi\pi}^s(z,\omega^2)&=&\frac{3F_{\pi\pi}^{\perp}(\omega^2)}{2\sqrt{2N_c}}(1-2z)\left[1
+a^s_{2\rho}(10z^2-10z+1)\right]  \;,\\
\phi_{\pi\pi}^t(z,\omega^2)&=&\frac{3F_{\pi\pi}^{\perp}(\omega^2)}{2\sqrt{2N_c}}(1-2z)^2\left[1
+a^t_{2\rho}\frac{3}{2}(5(1-2z)^2-1)\right]  \;,\\
\phi_{\pi\pi}^T(z,\omega^2)&=&\frac{3F_{\pi\pi}^{\perp}(\omega^2)}
{\sqrt{2N_c}}z(1-z)[1+a^{T}_{2\rho}\frac{3}{2}(5(1-2z)^2-1)]\;,\\
\phi_{\pi\pi}^a(z,\omega^2)&=&\frac{3F_{\pi\pi}^{\parallel}(\omega^2)}
{4\sqrt{2N_c}}(1-2z)[1+a_{2\rho}^a(10z^2-10z+1)]\;,\\
\phi_{\pi\pi}^v(z,\omega^2)&=&\frac{3F_{\pi\pi}^{\parallel}(\omega^2)}
{8\sqrt{2N_c}}\bigg\{[1+(1-2z)^2]+a^v_{2\rho}[3(2z-1)^2-1]\bigg\}\;,\\
\phi_{KK}^0(z,\omega^2)&=&\frac{3F_{KK}^{\parallel}(\omega^2)}{\sqrt{2N_c}}z(1-z)\left[1
+a^0_{2\phi}\frac{3}{2}(5(1-2z)^2-1)\right]\label{eq:phi0} \;,\\
\phi_{KK}^s(z,\omega^2)&=&\frac{3F_{KK}^{\perp}(\omega^2)}{2\sqrt{2N_c}}(1-2z)\;,\\
\phi_{KK}^t(z,\omega^2)&=&\frac{3F_{KK}^{\perp}(\omega^2)}{2\sqrt{2N_c}}(1-2z)^2 \;,\\
\phi_{KK}^T(z,\omega^2)&=&\frac{3F_{KK}^{\perp}(\omega^2)}
{\sqrt{2N_c}}z(1-z)[1+a^{T}_{2\phi}\frac{3}{2}(5(1-2z)^2-1)]\;,\\
\label{eq:phiT}
\phi_{KK}^a(z,\omega^2)&=&\frac{3F_{KK}^{\parallel}(\omega^2)}
{4\sqrt{2N_c}}(1-2z)\;,\\
\phi_{KK}^v(z,\omega^2)&=&\frac{3F_{KK}^{\parallel}(\omega^2)}
{8\sqrt{2N_c}}[1+(1-2z)^2]\;\label{eq:phiv},
\end{eqnarray}
with the $P$-wave form factors $F_{\pi\pi}^{\parallel(\perp)}(\omega^2)$ and $F_{KK}^{\parallel(\perp)}(\omega^2)$.
The values of the Gegenbauer moments associated with longitudinal and transverse polarization components are adopted the same as those  in Refs.~\cite{prd105-093001,prd104-096014,prd98-113003,epjc83-974}:
\begin{eqnarray}\label{eq:gen}
a^0_{2\phi}&=&0.40\pm 0.06, \quad a^T_{2\phi}=1.48\pm0.07,\non
a^0_{2\rho}&=&0.39\pm0.11, \quad  a^s_{2\rho}=-0.34\pm0.26,\quad a^t_{2\rho}=-0.13\pm0.04, \non
a^{T}_{2\rho}&=&0.50\pm0.50, \quad a^a_{2\rho}=0.40\pm0.40,\quad\quad a^v_{2\rho}=-0.50\pm0.50.
\end{eqnarray}
Because the amounts of the current experimental data are not yet enough for fixing the Gegenbauer moments in the twist-3 DAs $\phi_{KK}^{s,t}$ and $\phi_{KK}^{v,a}$,
they have been set to the asymptotic forms in our work.

In the experimental analysis of the multi-body hadronic $B$ meson decays,
the contribution from the wide $\rho$ resonant is usually parameterized as the Gounaris-Sakurai (GS)
model~\cite{prl21-244} based on the BW function~\cite{BW-model}.
By taking the $\rho-\omega$ interference and the excited states into account,
the form factor $F_{\pi\pi}^{\parallel}(\omega^2)$ can be written in the form of~\cite{prd86-032013}
\begin{eqnarray}
F_{\pi\pi}^{\parallel}(\omega^2)= \left [ {\rm GS}_\rho(s,m_{\rho},\Gamma_{\rho})
\frac{1+c_{\omega} {\rm BW}_{\omega}(s,m_{\omega},\Gamma_{\omega})}{1+c_{\omega}}
+\sum\limits_i c_i {\rm GS}_i(s,m_i,\Gamma_i)\right] \left[ 1+\sum\limits_i c_i\right]^{-1}\;,
\label{GS}
\end{eqnarray}
where $s=\omega^2$ is the two-pion invariant mass squared, $i=(\rho^{\prime}(1450), \rho^{\prime \prime}(1700), \rho^{\prime \prime \prime}(2254))$,
$\Gamma_{\rho,\omega,i}$ is the decay width for the relevant resonance, $m_{\rho,\omega,i}$ are the masses of the corresponding mesons, respectively.
The explicit expressions of the function ${\rm GS}_\rho(s,m_{\rho},\Gamma_{\rho})$ can be written as~\cite{BW-model}
\begin{equation}
{\rm GS}_\rho(s, m_\rho, \Gamma_\rho) =
\frac{m_\rho^2 [ 1 + d(m_\rho) \Gamma_\rho/m_\rho ] }{m_\rho^2 - s + f(s, m_\rho, \Gamma_\rho)
- i m_\rho \Gamma (s, m_\rho, \Gamma_\rho)}~,
\end{equation}
with the factors
\begin{eqnarray}
d(m) &=& \frac{3}{\pi} \frac{m_\pi^2}{k^2(m^2)} \ln \left( \frac{m+2 k(m^2)}{2 m_\pi} \right)
   + \frac{m}{2\pi  k(m^2)}
   - \frac{m_\pi^2  m}{\pi k^3(m^2)}~,\non
f(s, m, \Gamma) &=& \frac{\Gamma  m^2}{k^3(m^2)} \left[ k^2(s) [ h(s)-h(m^2) ]
+ (m^2-s) k^2(m^2)  h'(m^2)\right]~,\non
\Gamma (s, m_\rho, \Gamma_\rho) &=& \Gamma_\rho  \frac{s}{m_\rho^2}
\left( \frac{\beta_\pi (s) }{ \beta_\pi (m_\rho^2) } \right) ^3~.
\end{eqnarray}
The functions $k(s), h(s)$ and $\beta_\pi (s)$ can be expressed as
\begin{eqnarray}
k(s) =\frac{1}{2} \sqrt{s}  \beta_\pi (s)~, \quad
h(s) = \frac{2}{\pi}  \frac{k(s)}{\sqrt{s}}  \ln \left( \frac{\sqrt{s}+2 k(s)}{2 m_\pi} \right),\quad
\beta_\pi (s) = \sqrt{1 - 4m_\pi^2/s}.
\end{eqnarray}

For the vector form factor of the $K{\bar K}$ system,
the dominant resonance is $\phi(1020)$ in the concerned mass window.
We then employ the relativistic BW line shape to parameterize the $F_{KK}^{\parallel}(\omega^2)$~\cite{epjc78-1019},
\begin{eqnarray}
\label{BRW}
F_{KK}^{\parallel}(\omega^2)&=&\frac{ m_{\phi}^2}{m^2_{\phi} -\omega^2-im_{\phi}\Gamma_{\phi}(\omega^2)} \;,
\end{eqnarray}
with the mass-dependent width $\Gamma_{\phi}(\omega^2)$
\begin{eqnarray}
\label{BRWl}
\Gamma_{\phi}(\omega^2)&=&\Gamma_{\phi}\left(\frac{m_{\phi}}{\omega}\right)\left(\frac{k(\omega)}{k(m_{\phi})}\right)^{(2L_R+1)}.
\end{eqnarray}
The $m_{\phi}=1.0195$ GeV~\cite{pdg2020} and $\Gamma_{\phi}=4.25$ MeV~\cite{pdg2020} represent the mass and natural width  of the $\phi$ meson, respectively.
The orbital angular momentum $L_R$ in the two-meson system is set to $L_R = 1$ for a $P$-wave state.
The $k(\omega)$ is the momentum vector of the resonance decay product measured in the resonance rest frame, while $k(m_{\phi})$
is the value of $k(\omega)$ when $\omega=m_{\phi}$.
Due to the limited studies on the form factor $F^{\perp}(\omega^2)$,
we usually assume the approximation $F^{\perp}(\omega^2)/F^{\parallel}(\omega^2)\approx f_{V}^T/f_{V}$ in our calculation,
with $f_{V}^{(T)}$ being the vector (tensor) decay constants of the intermediate vector resonance.


\end{document}